\documentclass[a4paper,11pt]{article}
\pdfoutput=1 
\usepackage{jcappub} 
\usepackage[T1]{fontenc} 
\usepackage[dvipsnames]{xcolor}
\usepackage{ragged2e}
\usepackage{mathtools}

\usepackage{subcaption}
\usepackage{float}
\usepackage{amsmath}
\usepackage{array}
\usepackage{ragged2e}
\usepackage{hyperref}
\hypersetup{
    colorlinks=true,
    linkcolor=blue,
    filecolor=magenta,
    urlcolor=blue,
    citecolor=magenta
}
\usepackage{color}
\usepackage{url}
\usepackage{multirow, makecell}

\usepackage{graphicx}
\usepackage{float}
\usepackage[nodisplayskipstretch]{setspace}
\newcolumntype{C}{ >{\centering\arraybackslash} m{3cm} }
\newcolumntype{D}{ >{\centering\arraybackslash} m{4.5cm} }
\newcolumntype{E}{ >{\centering\arraybackslash} m{6cm} }

\newcommand{\zobs}{z_{\textrm{obs}}}
\newcommand{\zp}{Z^\prime}

\usepackage{tikz}
\usetikzlibrary{fit,shapes.misc}

\title{\boldmath Deep learning techniques for Imaging Air Cherenkov Telescopes}

\author[a]{Songshaptak De,}
\author[a]{Writasree Maitra,}
\author[a]{Vikram Rentala}
\author[b]{and Arun M. Thalapillil}

\affiliation[a]{Department of Physics, Indian Institute of Technology Bombay, Powai, Mumbai, Maharashtra, 400076, India}
\affiliation[b]{Department of Physics, Indian Institute of Science Education and Research Pune, Pune, Maharashtra, 411008, India}

\emailAdd{saptak@phy.iitb.ac.in}
\emailAdd{writasreemaitra@iitb.ac.in}
\emailAdd{rentala@phy.iitb.ac.in}
\emailAdd{thalapillil@iiserpune.ac.in}

\justifying

\abstract{
Very High Energy (VHE) gamma rays and charged cosmic rays (CCRs) provide an observational window into the acceleration mechanisms of extreme astrophysical environments. One of the major challenges at Imaging Air Cherenkov telescopes (IACTs) designed to look for VHE gamma rays, is the separation of air showers initiated by CCRs which form a background to gamma ray searches. Two other less well-studied problems at IACTs are a)~the classification of different primary nuclei among the CCR events and b)~identification of anomalous events initiated by Beyond Standard Model (BSM) particles that could give rise to shower signatures which differ from the standard images of either gamma rays or CCR showers. The problems of categorizing the primary particle that initiates a shower image, or the problem of tagging anomalous shower events in a model independent way, are problems that are well suited to a machine learning (ML) approach. Traditional studies that have explored gamma ray/CCR separation have used a multivariate analysis based on derived shower properties, which contains significantly reduced information about the shower. In our work, we address the problems outlined above by using machine learning architectures trained on full simulated shower images, as opposed to training on just a few derived shower properties. We illustrate the techniques of binary and multi-category classification using convolutional neural networks, and we also pioneer the use of autoencoders for anomaly detection at VHE gamma ray experiments. The latter technique has been studied previously in the context of collider physics, to tag anomalous BSM candidates in a model-independent way. In this study, for the first time, we demonstrate the efficacy of these techniques in the domain of VHE gamma ray experiments. As a case study, we apply our techniques to the H.E.S.S. experiment. However, the real strength of the techniques that we broach here in the context of VHE gamma ray observatories, is that these methods can be applied broadly to any other IACT --- such as the upcoming Cherenkov Telescope Array (CTA) --- or can even be suitably adapted to CCR experiments.}

\makeatletter
\gdef\@fpheader{}
\makeatother

\begin{document}
\maketitle
\flushbottom

\justify

\section{Introduction}
\label{sec:intro}
Very High Energy (VHE) gamma rays are photons with energies between $\sim$~100~GeV to 100~TeV that bombard the Earth's atmosphere~\cite{2012EPJH...37..459L,2021ChA&A..45..281S}. Charged cosmic rays (CCRs) on the other hand are primarily made up of protons and alpha particles, with a small admixture of heavier charged nuclei, electrons, and anti-particles ($\overline{p},e^+$ etc.)~\cite{doi:10.1146/annurev.ns.33.120183.001543}. Charged cosmic rays with energies between 30 GeV and 3-4 PeV have a nearly power law spectrum, with a spectral index given by $\Gamma = - 2.7$  up until the so called ``knee'' at $\sim$~4~PeV~\cite{Castellina:2011gn}. These charged cosmic rays are expected to be of galactic origin, and are speculated to originate from supernovae remnant shocks~\cite{Amato:2014xua,Bykov:2018wrt}, although it remains an open question what the actual sources of PeV CCRs are~\cite{Cristofari:2021jkl}.

The study of VHE gamma rays and CCRs can help us understand the origins of the extreme astrophysics that lead to acceleration of particles to such high energies~\cite{TeVCat,Bose:2022ghr}. Historically, cosmic rays have also played a significant role in the discovery of new physics and particles, such as muons and strange quarks~\cite{doi:10.1063/1.4792534,Friedlander:2012zz}. This is because they provide a natural particle accelerator, allowing access to energies well beyond the reach of terrestrial collider experiments. It has also been speculated that high energy gamma rays, or charged cosmic rays could arise from the annihilation of dark matter in our galaxy~\cite{Leane:2020liq,Slatyer:2021qgc} and thus characterizing the spectrum and spatial distribution of cosmic rays could help pin down such exotic origins~\cite{2021JCAP...01..057A}.

VHE gamma rays as well as charged cosmic rays in the energy range of tens of GeV to 100 TeV can be detected at imaging air Cherenkov telescopes (IACTs)~\cite{Krennrich_2009,KNODLSEDER2016663}. The basic principle is that when a gamma ray (or charged cosmic ray) strikes particles in the upper atmosphere, it produces an ~``extensive air shower'' (EAS) which is a cascade of particles that in turn emit Cherenkov radiation. Ground based telescopes can be used to detect the Cherenkov radiation. The shower image seen in the telescopes can be used to reconstruct the energy, angle, and type of primary particle that initiated the cosmic ray shower.

Three current generation IACTs are successfully collecting data for VHE gamma rays with energies between 30 GeV - 100 TeV at the present time - H.E.S.S.~\cite{Bernlohr:2003tfz,Cornils:2003ve}~in Namibia; VERITAS~\cite{Weekes:2001pd} in Arizona and MAGIC~\cite{Cortina:2004qt} in La Palma.  A more sophisticated and highly sensitive next-generation IACT array, the Cherenkov Telescope Array (CTA)~\cite{2013APh....43....3A}, is being built which will have arrays of more than 100 telescopes situated across two sites, one each in the northern and southern hemispheres. CTA will be sensitive to gamma rays between 20~GeV and 300~TeV.

Both VHE gamma rays and CCR showers can produce Cherenkov light which can be detected at IACTs. However, VHE gamma rays typically lead to relatively narrow electromagnetic showers (containing mostly $e^-$, $e^+$, and $\gamma$), whereas CCRs produce broader showers, with additional components such as additional hadrons, $\pi, K$-mesons and their decay products $\gamma,\mu^\pm, \nu$. Dedicated ground based detectors for CCRs also attempt to detect these extra particles in the cascade, for example by using muon detectors. CCRs are typically assumed to form a background to searches of gamma rays at IACTs, and separation of CCR hadronic images from gamma ray initiated shower images is a critical task at these experiments.

Conventionally, the strategy used to separate CCR images at IACTs relies on the fact that CCR showers are typically wider than gamma ray initiated showers. The shower images are fit to the so called Hillas parameters~\cite{1985ICRC....3..445H}, which characterize the elliptical shape of the shower image. From simulation, the range of Hillas parameters that are expected from CCRs or gammas with a particular energy and angle are known, and this can be used to infer properties of the cosmic ray primary at an IACT experiment.

While CCRs have long been regarded as a background at IACTs designed to look for gamma rays, it is also possible to attempt to detect CCRs and characterize the spectrum and composition of different primary species at these experiments. While this problem has received little attention in the literature, a recent effort in this direction was made at the H.E.S.S. experiment~\cite{Jankowsky:2020baz}\footnote{Another study along similar lines was performed in ref.~\cite{2018PhRvD..98f2004A}, which looked into measurement of the cosmic-ray electron and positron spectrum at VERITAS.}.

Another intriguing (and less studied) possibility is that events that are rejected as background at IACT experiments may contain signals of new beyond Standard Model (BSM) physics. New BSM signatures could result from either collisions of Standard Model primaries with the atmosphere, or they could be produced through the decay of dark matter particles or other exotica in the upper atmosphere\footnote{The energy range of CCRs from 100 GeV - 100 TeV, leads to a center of mass collision energy of 10 GeV - 300 GeV which is less than the center-of-mass energy at the Large Hadron Collider (LHC). Moreover the low CCR fluxes would lead to a far lower event rate at IACTs than at the LHC. Thus, it is unlikely that any exotica would be produced in CCR collisions given that exotic particle physics at these center-of-mass energies is strongly constrained at the LHC. However, the alternative possibility of decaying dark matter particles or other exotica remains an open possibility. For the purposes of our work, we remain agnostic as to the source of the exotic BSM physics.}.

In this work, we are interested in exploring the following three problems of interest at VHE gamma ray IACT experiments:

\begin{enumerate}
    \item \textbf{Binary classification:} Can we predict whether IACT images are initiated by a particular SM primary (such as a high energy gamma) or by some other particles? This would correspond to the typical problem of gamma-hadron separation at an IACT.
    \item \textbf{Multi-category classification:} Going further than binary classification, can we correctly identify or categorize images based on the particular Standard Model primary that initiated the shower? This would correspond to attempting to identify the specific species of CCR primary, or gamma ray that initiated a shower image.
    \item \textbf{Anomaly detection:} Can we flag potentially anomalous events that have features that do not conform to ``standard'' images expected for showers initiated by SM primaries? This would correspond to identifying potential BSM candidate events at IACTs. For similar applications at charged cosmic ray experiments, see for example~~\cite{Schichtel:2019hfn,Reininghaus_2021}.
\end{enumerate}

Machine learning (ML) techniques are perfectly suited to address the problems listed above.  In the present work, we shall focus on two main classes of ML algorithms: supervised learning and unsupervised learning~\cite{10.5555/1162264}.

In supervised learning, known templates (in our case shower images), can be fed to a machine (such as a neural net) with labels specifying the category of the image (in this case specifying the primary particle that initiated the shower image). The machine then learns essential features of the image that correspond to a given category. This trained machine can then be deployed in the field on images that it has not previously encountered in order to determine which category they correspond to. Binary and multi-category classification of cosmic ray primaries\footnote{Henceforth in this paper, whenever the term ``cosmic ray'' shower appears, it will generically refer to showers initiated either by gamma rays or by CCRs unless otherwise stated.} are problems for which supervised learning algorithms may be well adapted.

Anomaly detection on the other hand is a problem for which unsupervised machine learning is better suited. In this case, we train our machine on standard model images without labels, and the machine learns essential features of these training images. Then when deploying the machine on new classes of images (such as an image initiated by an exotic BSM $Z^\prime$\footnote{There are a vast number of models for such BSM Z$^\prime$ particles. Readers interested in details of some of these models can see for e.g.~\cite{DelAguila:1993px,Langacker:2008yv,Hayden:2013sra}.}), the machine can flag such events as anomalous because they do not conform to the standard features that the machine has learnt.

ML thus provides us with a powerful set of tools to address the typical problems faced at IACTs. While conventional ML has been used at IACTs for some of the purposes described above (most notably for gamma-hadron separation), historically the tools have mostly been used in a limited way, for example using Hillas parameters (possibly in conjunction with a few additional discrimination variables) as inputs to a Random Forest~\cite{Albert_2008} or a Boosted Decision Tree (BDT)~\cite{Ohm_2009, Becherini_2011, 2017APh....89....1K}, which results in an analysis which has a level of sophistication similar to that of a typical multi-variate analysis.

With advances in computational power, more recent studies have attempted to leverage advanced ML techniques such as Deep Neural Networks (DNNs) for background rejection (gamma-hadron separation) at IACTs~\cite{2017ICRC...35..809N,2018arXiv181000592M,2018arXiv181201551P,Brill_2019,2019ASPC..523...75A,2019APh...105...44S,2019EPJWC.21406020V,2020JPhCS1525a2084L,2021arXiv210107626N,2021APh...12602527B}. Some of these studies have also tried to leverage DNNs to reconstruct other shower properties as well, such as the energy and angle of the primary particle~\cite{2021arXiv210804130V,  Aschersleben_2021}. DNNs are a type of Artificial Neural Network that possess many layers which allow them to extract complex features of a raw input data set in a highly efficient manner. The subtype of DNNs that are most efficient for image analysis are Convolutional Neural Networks (CNNs). Several packages implementing such networks have been created for use at IACTs~\cite{2021arXiv210107626N,2019JPhCS1181a2048P, 2021arXiv210905809M}.
These works rely on CNNs which learn features of input simulation telescopic images used for training. These CNNs can then be used to reconstruct the same properties of the shower in actual data.

One of the challenges involved with using IACT images as inputs to a DNN is the non-square nature of the telescopic image~\cite{2019arXiv191209898N,2019SoftX...9..193S}, which needs to be either reshaped or padded into a square array, or alternatively the input structure to the CNN needs to be modified to preserve topological information about the input image. A related issue is that a single event can show up in multiple telescopes and information from all the telescopes has to be collated and treated as a single input to the machine~\cite{Brill_2019}. Additionally, with more sophisticated telescopes, one may attempt to use not only static telescopic images for each event, but also the time series wave forms for gamma-hadron separation~\cite{2019ICRC...36..798S,2021APh...12902579S}.

In this work, we seek to leverage the full power of machine learning by applying CNNs directly to the full telescopic shower images seen at IACTs. This kind of learning utilizes the full information collected by the detector and thus can be sensitive to more features than just the reconstructed Hillas parameters.

In the context of VHE gamma ray observatories and exotic BSM event tagging at these experiments, we for the first time in the literature are broaching the use of CNNs and autoencoders~\cite{10.5555/3153997} --- to learn complex features of the respective image categories and in addition, for the latter approach, facilitating the flagging of anomalous BSM images in a model independent way. Such techniques with autoencoders, for anomaly detection, have been proposed for use at collider physics experiments like the LHC~\cite{Heimel_2019, Farina_2020, roy2020robust,finke2021autoencoders,aarrestad2021dark,Collins_2021}, however they had hitherto not fully made their way to cosmic ray/VHE gamma ray experiments\footnote{See \cite{2021arXiv210700656B} for a recent proposal to develop cross-disciplinary machine learning tools for applications in fundamental physics. The portability of these tools across different types of experiments is especially relevant in the context of broad model-independent search strategies.}.

Our work is intended to be a proof-of-concept study on the applicability of these advanced machine learning techniques to gamma ray/CCR experiments, and we hope that future studies will build upon these ideas in order to maximally leverage the data being collected at these experiments.

Although our study is presented in the context of IACTs, since our methods are based on shower property reconstruction from detector event images, we expect that our basic techniques can be easily adapted to solve similar problems of CCR primary classification and BSM primary identification at CCR experiments such as the Pierre Auger Observatory~\cite{PierreAuger:2015eyc}, HAWC~\cite{SPRINGER201687}, and LHASSO~\cite{DiSciascio:2016rgi}.

In the next section, we will describe the basic strategy that we will follow, and then we will explain the outline of the paper in light of  this strategy.

\section{Description of the overall strategy }
\label{sec:strategy}

Our strategy will be as follows:
\begin{enumerate}
\item First we will generate ``standard'' telescopic images of showers generated by gamma rays and CCR species (proton,  helium, carbon). These images are then combined into a single JPEG image which shows the responses of all the telescopes. As a test case, the shower images are generated for the H.E.S.S. telescope.
\item We also generate ``anomalous'' images of a $Z^\prime$ going to  an electron-positron pair as an example of a prototypical BSM event. Since we are concerned with the ability of our machine learning algorithms to flag such anomalous BSM events in a model agnostic way, we leave origin of the $Z^\prime$ unspecified.

\item With the standard images, we build a bank of labelled images corresponding to gamma and CCR primaries. These images are used for supervised learning for binary and multi-category classification as labelled data sets, with labels corresponding to the primary particle type. For tagging anomalous events, we use unsupervised learning, where our autoencoder is trained on standard images, where no labels (information about the type of primary) are passed to the autoencoder.
\item Supervised learning: We train a binary/multi-category classifier on our labelled images and test the performance of the machine based on how well it can separate different categories of images. This classification method can be used for gamma/hadron separation or alternatively, to identify different hadronic CCR species at a gamma ray detector.
\item Unsupervised learning: We train our autoencoder using standard images and see how often it is able to flag anomalous events that it has not encountered in training and that do not conform to features of the standard images that it has previously learned. Here we pass $Z^\prime$ images for testing to the autoencoder, but the autoencoder should in principle work for any type of anomalous images.
\item For each objective, either binary/multi-category classification, or anomaly detection, we need to define some figures-of-merit to gauge the performance of our machines. These figures-of-merit are computed on test simulation data that the neural nets have not previously encountered during the training phase. These numbers quantify the performance of our machines.
\end{enumerate}

Based on the above strategy, we now explain the outline of our paper.
In section \ref{sec:showerimage} we describe the details of the shower simulation and image generation. Then in section \ref{sec:architecture} we describe the setup of our machine learning architectures for supervised learning (for binary and multi-category classification) and unsupervised learning (anomaly detection). We will also describe various figures-of-merit for characterizing the performance of these machines in this section. In section~~\ref{sec:Results}, we present the performance results of our classifiers and autoencoder by computing the figures-of-merit for both the supervised and unsupervised machine learning techniques. We summarize and discuss some of the advantages as well as limitations of our method along with possible applications and extensions of our techniques in section~\ref{sec:Conclusions}. In the appendices, we present our results for some other combinations of cosmic ray primary energies and zenith angles to validate the robustness of our results.

\section{Generating shower images for different categories of events}
\label{sec:showerimage}

The High Energy Stereoscopic System (H.E.S.S.) detector is an array of IACTs located in Namibia at a height of 1800 m above sea-level~\cite{Bernlohr:2003tfz,Cornils:2003ve}. The H.E.S.S. Phase-I telescopic system consists of four telescopes which are arranged at the corners of a square of side 120~m\footnote{There is a fifth telescope, HESS-II, which is much larger in size in comparison to the four phase-I telescopes and is located at the center of the array of the four phase-I telescopes. In our work, we have not simulated this fifth telescope.}. Each H.E.S.S. Phase-I telescope has a dish of diameter 12~m containing 382 circular mirrors. Each dish has four arms which support a camera placed above the center of the dish. The camera is placed at the focal point of the dish, at a distance of 15~m. The camera has 960 hexagonal pixels  which we will refer to as ``detector pixels'' henceforth (to distinguish them from pixels of the RGB images that we will use later to represent the full detector image). The pixels in each camera are arranged in the form of an octagonal lattice. Each telescope has a field-of-view of 5$^\circ$. The whole telescopic dish structure with the camera can be rotated, both in the horizontal and the vertical plane.

We will assume for simplicity that all the telescopes are aligned to point in the same direction. Thus, to determine a specific configuration of the H.E.S.S. telescope array, one needs to specify the angle of rotation in the horizontal plane (the azimuthal angle, $\phi_{\textrm{tel}}$ with respect to some reference direction) and an angle in the vertical plane (the zenith angle, $\theta_{\textrm{tel}}$). The arrangement of the telescopic system and these angles are shown in Fig.~\ref{fig:tel}.

Now, let us consider the physics leading to the formation of the images in the H.E.S.S. telescopic system. These images with all the telescopic effects incorporated will be the main inputs to our machine learning architectures. Relativistic charged particles produced in a cosmic ray shower produce Cherenkov light as they traverse through the atmosphere. The Cherenkov light from a typical shower projects onto an elliptical region on the ground. This is called a Cherenkov light pool. The H.E.S.S. telescopes are designed to capture these Cherenkov light pools and digitize them into detector pixel intensities (photoelectron counts) as seen in their cameras.

The images seen by all four telescopes, from their respective vantage points, can be collected into a single image that contains all the observed data for a single cosmic ray event. In the rest of this section, we will describe in detail the procedure we followed to generate such images for our pool of standard and anomalous events.

Our simulation and extraction methodology consists of three stages,
\begin{itemize}
\item \textbf{Extened Air Shower (EAS) Simulation} - We first simulate the EAS initiated by a standard set of SM particles - gamma ($\gamma$), proton ($p$), helium (He) and carbon (C). This choice is  motivated by what the dominant primary particles contributing to cosmic rays at our simulated energies are. It may in principle be enlarged to include more nuclei. This simulation is done using \texttt{CORSIKA}~\cite{CorsikaUserguide}.
\item \textbf{Telescopic Simulation} - Next we simulate detector effects in the H.E.S.S. telescopes, which gather the Cherenkov light pools generated in the shower. These detector effect simulations are performed with the aid of \texttt{sim\_telarray}~\cite{Bernl_hr_2008}.
\item \textbf{RGB Image Extraction} - In the final phase, we convert all the individual H.E.S.S. telescopic pixel intensities into a single composite image, which shows all the four individual telescopes, with their relative pixel locations and intensities. For this image generation we will make use of the \texttt{ctapipe} package~\cite{ctapipe}.
\end{itemize}

We now describe our procedure for simulation and image generation following the three steps outlined above. We will first describe the methodologies for the standard particles. Later, in Sec.~\ref{sec:anomalysim}, we will describe how we extend these to the anomalous $Z^\prime$ images. As we mentioned, the $Z^\prime$ initiated shower will be the prototypical new physics anomalous event in our study. To faithfully simulate these, we will see in Sec.~\ref{sec:anomalysim} how the standard simulation procedure has to be slightly modified to overcome the limitations of the simulation software that we are using when simulating BSM primaries.

\subsection{Standard events}
\label{sec:standard_sim}

\subsubsection{Extended air shower simulation}
\label{sec:EAS_sim}

To simulate the EAS from our standard set, we use the air shower simulator, \texttt{CORSIKA} (COsmic Ray SImulations for KAscade) \cite{CorsikaUserguide}. Within \texttt{CORSIKA} we use \texttt{QGSJET01} and \texttt{GHEISHA} as the high energy and low energy hadronic interaction models, respectively. All runtime options for a shower simulation in \texttt{CORSIKA} are configured in an `input card'. The main inputs that we have selected are i) the properties of the primary particles (particle type, energy,  direction, and starting height), ii) the  telescopic array description (this sets up the coordinate positions of the telescopes, which is required to simulate the recording of Cherenkov photons in a spherical region around each telescopic location) iii) the local magnetic field, iv) the atmospheric profile. \texttt{CORSIKA} simulates the EAS -- based on the primary properties, the atmospheric profile, and the local magnetic field -- and then records the Cherenkov radiation that can potentially be seen by the telescopes.

The last three of the inputs listed above are set by the specific details of the H.E.S.S. experiment. For example, we take into account the atmospheric profile that matches that of the H.E.S.S. location in Namibia. We also take in account the geomagnetic declination for H.E.S.S. site in the air shower simulation.

\begin{figure}
\centering
\includegraphics[height=9cm,width=10cm]{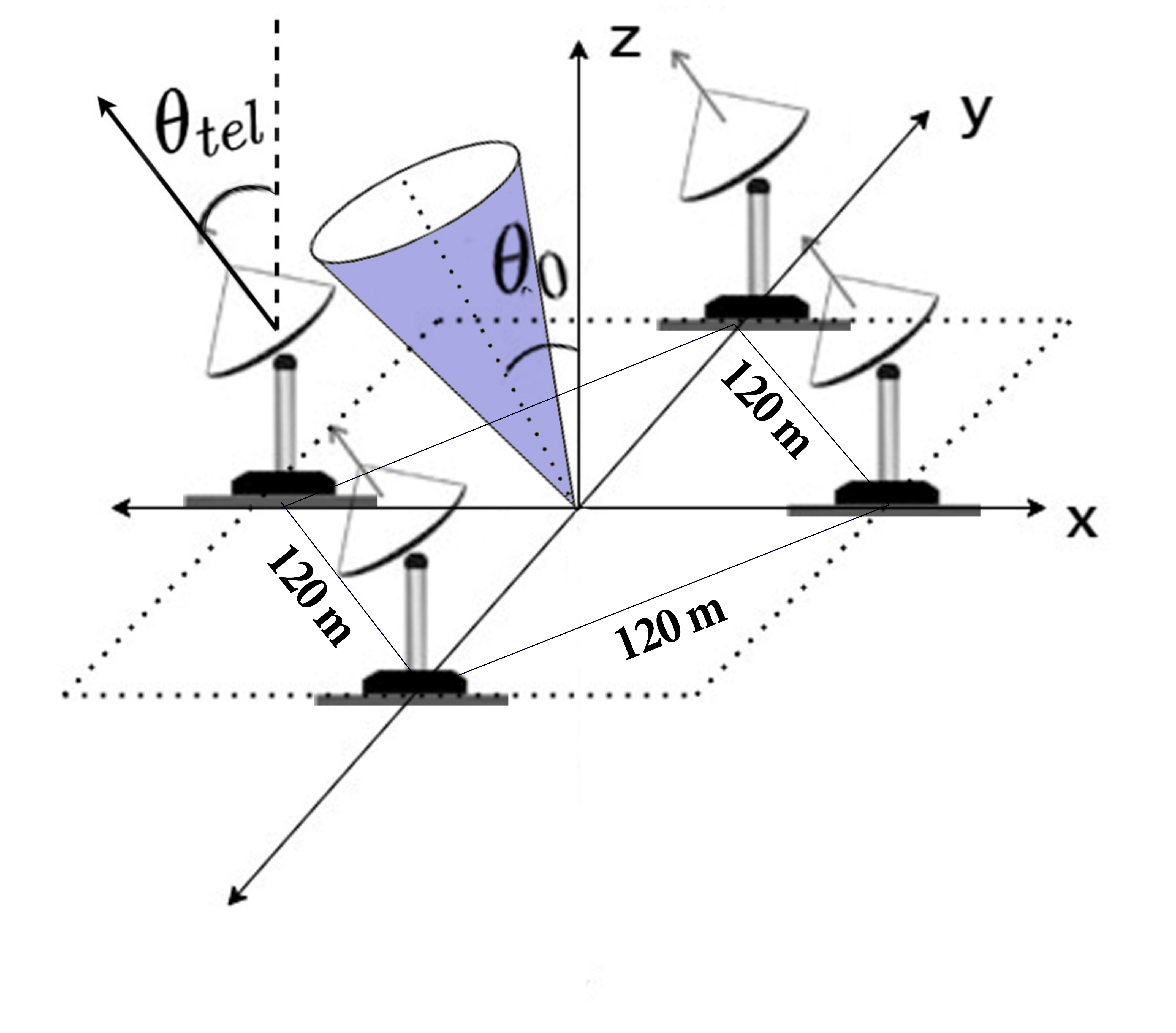}
\caption{Pictorial representation of the four H.E.S.S. Phase-I telescopes showing their spatial arrangement at the corners of a square of side 120~m in the observation plane. The zenith pointing angle $\theta_{\textrm{tel}}$ is also shown. The cosmic ray shower axis is chosen to lie in a cone (depicted in blue) with vertex at the detector center, and with semi-vertical angle $1.5^\circ$, where the cone-axis has zenith and azimuthal angles $\theta_0$ and $\phi_0$, respectively. In our analyses, we will always take the detector pointing direction to be in the same direction as the direction of the cone axis.}
\label{fig:tel}
\end{figure}

The input for the H.E.S.S. telescopic configuration in \texttt{CORSIKA} specifies four telescopes at the corners of a square of side 120~m. In \texttt{CORSIKA}'s co-ordinate system, the $(x,y,z)$ co-ordinates $(0,0,\zobs)$ (where $\zobs=1800$~m is the observation height above sea-level) by default corresponds to the point at which the shower axis intersects the detector plane.
For simplicity, we choose the center of the telescopic system to be at this point where the shower axis intersects the detector plane, i.e. all the standard simulated events are those where the primary particle is headed for the center of the telescopic system, for any choice of primary particle incident direction ($\theta_{\textrm{shower}}$ and $\phi_{\textrm{shower}}$).

For simulation of the standard events, the primary particle types are specified by fixing the \texttt{CORSIKA} particle IDs\footnote{We note in passing that the \texttt{CORSIKA} particle IDs are different from the PDG particle IDs.} corresponding to $\gamma$, proton, carbon and helium. For our main analysis, we will generate EAS showers by selecting primary particles with energies between $E_{\textrm{min}}=100~-~0.5$~TeV and $E_{\textrm{max}}=100~+~0.5$~TeV. The energies are randomly generated in this range. This is done by sampling from a power law distribution with a probability density function given by,
\begin{equation}
P(E)=
\begin{cases}
\frac{\Gamma+1}{E_{\textrm{max}}^{\Gamma+1}-E_{\textrm{min}}^{\Gamma+1}}E^\Gamma \qquad \text{where},~~ E_{\textrm{min}}<E<E_{\textrm{max}}, \\[4mm]
0 \qquad \qquad \qquad \qquad \text{otherwise,}
\end{cases}
\end{equation}
where we have taken $\Gamma=-2.7$ to be a representative spectral index for all our cosmic ray primaries. We will discuss the results of our analysis with other choices of energy in the appendices.

In order for the H.E.S.S. telescope array to see most of the cosmic ray shower, the primary particles must be travelling approximately along the viewing direction of the telescopes. We thus randomly select the direction of the shower (parameterized by $\theta_{\textrm{shower}}$ and $\phi_{\textrm{shower}}$) to lie within a cone of semi-vertical angle $1.5^\circ$, with vertex fixed at the telescopic center, and with the direction of the cone-axis fixed at some $\theta_0$ and $\phi_0$ which need to be specified in \texttt{CORSIKA}. For our main analysis, we will use $\theta_0=0^\circ$ and $\phi_0=0^\circ$. With this choice, the cone axis is perpendicular to the ground, i.e. cosmic rays headed down the cone axis would be coming straight down.  We will also show our results for other choices of these angles in the appendices. The semi-vertical angle of this cone is chosen keeping the field of view (FOV) of the H.E.S.S. telescope ($5^\circ$ FOV) in mind. In \texttt{CORSIKA} this selection is enabled by selecting the \texttt{VIEWCONE} option.

Finally, we allow the height of first interaction for the SM primaries to be randomly determined by \texttt{CORSIKA}. We allow the primary to propagate from the top of the atmosphere to its first interaction point by selecting a starting grammage of 0 gm/cm$^2$ in the input card\footnote{Grammage is defined as the integrated column density seen by a cosmic ray along its propagation starting from the topmost point of the atmosphere. Thus, the grammage of the highest point of the atmosphere is 0~gm/cm$^2$.}.

When simulating an air shower for a very high energy primary, \texttt{CORSIKA}. attempts to generates a huge number of secondary particles which leads to long simulation times and large file sizes for the outputs. Sometimes, in order to bypass these issues, the THIN sampling method is opted for, which only retains a relevant subset of the secondary particles. For the energy range we are working with, the number of secondaries is not so large so as to warrant usage of the thinning option, so our simulations are performed without THIN sampling.

\subsubsection{Telescopic simulation}
\label{sec:simtel}

The next step of our simulation is to take the output of the air shower generated by \texttt{CORSIKA} and to pass this to \texttt{sim\_telarray}~\cite{Bernl_hr_2008}, to simulate the telescope response. The whole process of detector simulation in \texttt{sim\_telarray} mimics the propagation of Cherenkov photons from the air shower to the cameras placed at the center of each telescope, and the recording of the pixel intensities in digital format.

The input to \texttt{sim\_telarray} is the output file of \texttt{CORSIKA} that contains all the relevant information about the Cherenkov radiation of the EAS that can potentially be detected by the telescopes. The H.E.S.S. telescopic configuration is included by default in the \texttt{sim\_telarray} package. \texttt{sim\_telarray} simulates the telescopic response taking into account effects such as the dish shapes, roughness of the mirror surfaces, optical point spread functions, reflectivity, shadowing by the camera and its support structure, the angular acceptance of the pixels, and the quantum efficiency of photo-multiplier tubes. Night sky background effects are also incorporated in the telescopic simulation.

For our simulation, we fix our telescopic dish orientation by selecting the common zenith angle, $\theta_{\textrm{tel}}$ and azimuthal angle, $\phi_{\textrm{tel}}$ for all the four telescopes. We demonstrate a pictorial representation of how all the four H.E.S.S. phase I telescope dishes will orient themselves in Fig.~\ref{fig:tel}. Note that in general the telescopes might be pointing away from the shower axis. However, we will choose our telescopes to point in the direction of $\theta_0$ and $\phi_0$, so that, within the $1.5^\circ$ variability of the shower axis induced by the VIEWCONE option, the telescope viewing axis is aligned with the cosmic ray shower axis. For our main analysis with $\theta_0=0^\circ$ and $\phi_0=0^\circ$, our choice of $\theta_{\textrm{tel}}$ and $\phi_{\textrm{tel}}$ is such that the telescopes are pointing straight upwards.

In the H.E.S.S. telescope, an event is generally recorded if at least two of the telescopes are triggered\footnote{A camera trigger occurs if the signals in M pixels within a sector (sector threshold) exceed a threshold of N photoelectrons (pixel threshold). In H.E.S.S., $M = 3$ and $N = 5.3$ p.e.. This choice yields a trigger rate at H.E.S.S. of  $\mathcal{O}$(100) Hz~\cite{trig_pe}.}. In our study we will be more conservative and will only consider events where all four telescopes are triggered.

In Table~\ref{config_sim}, we summarize the different configuration options we select for our main analyses---including both the air shower simulation using \texttt{CORSIKA}, as well as the telescopic simulation.

\begin{table}[t]
\centering
\begin{tabular}{|C|D|E|}
\hline
 & & \\
Level & Configuration parameters & Configuration values  \\
 & & \\
\hline
 & &  \\
\multirow{5}{4em}{\texttt{CORSIKA}}&Spectral Index, $\gamma$ & -2.7  \\
& Energy range & ($100-0.5$) TeV to ($100+0.5$) TeV\\
& Zenith angle, $\theta_0$ & $0^\circ$\\
& Azimuthal angle, $\phi_0$ & $0^\circ$\\
& Viewcone angle &$1.5^\circ$\\
 & & \\
\hline
 & & \\
\multirow{2}{6em}{\texttt{sim\_telarray}}& Zenith angle, $\theta_{\textrm{tel}}$ & $0^\circ$\\
& Azimuthal angle, $\phi_{\textrm{tel}}$ & $166^\circ$\\
 & & \\
\hline
\end{tabular}
\caption{The different configuration parameters and their values used in both the EAS simulation using \texttt{CORSIKA} and in the telescopic simulation using \texttt{sim\_telarray}. The cosmic ray shower axis lies within a cone of semi-vertical angle $1.5^\circ$ with vertex fixed at the telescopic center and with the direction of the cone axis defined by the parameters $\theta_0$ and $\phi_0$ in \texttt{CORSIKA}. Note that with the reference axis conventions used in \texttt{sim\_telarray}, the choice of $\theta_{\textrm{tel}}$ and $\phi_{\textrm{tel}}$ are such that the telescopes point in the direction of the cone axis defined by the parameters $\theta_0$ and $\phi_0$ in \texttt{CORSIKA}. For the choices presented in the table, the telescope viewing direction and the cone axis are straight upwards. We consider other choices of angles in the appendix, but in all cases we will take the telescopes to point in the direction of the cone axis.}
\label{config_sim}
\end{table}

\subsubsection{RGB image extraction}
\label{sec:RGBimage}

The output file of \texttt{sim\_telarray} contains raw data of detector pixel intensities in ADCs (Analog to Digital Counts)\footnote{The H.E.S.S. detector has a high and a low gain channel. We are using the ADC output for the high gain channel only.}. The conversion of these ADCs into equivalent number of photoelectrons is an essential step for further analysis of these pixel intensities, whether it be for the reconstruction of the shower direction or the primary particle detection. This conversion is based on calibration of the H.E.S.S. telescopes.

After the detector pixel intensities are properly calibrated, they are cleaned to eliminate pixels that contain noise or night sky background photons. A two stage tail-cut procedure is followed for image cleaning. This means that detector pixels having amplitudes greater than $10$ photoelectrons, with boundary pixels of amplitude more than $5$ photoelectrons are accepted as is, but pixels not satisfying these constraints will be set to zero amplitude~\cite{cleaning, Hupfer:2008uta}.

\begin{figure}
\begin{center}
\begin{subfigure}{0.3\textwidth}
\includegraphics[height=4cm,width=4cm]{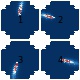}
\subcaption{}
\end{subfigure}
\begin{subfigure}{0.3\textwidth}
\includegraphics[height=4cm,width=4cm]{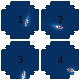}
\subcaption{}
\end{subfigure}
\begin{subfigure}{0.15\textwidth}
\includegraphics[height=3.8cm,width=6cm]{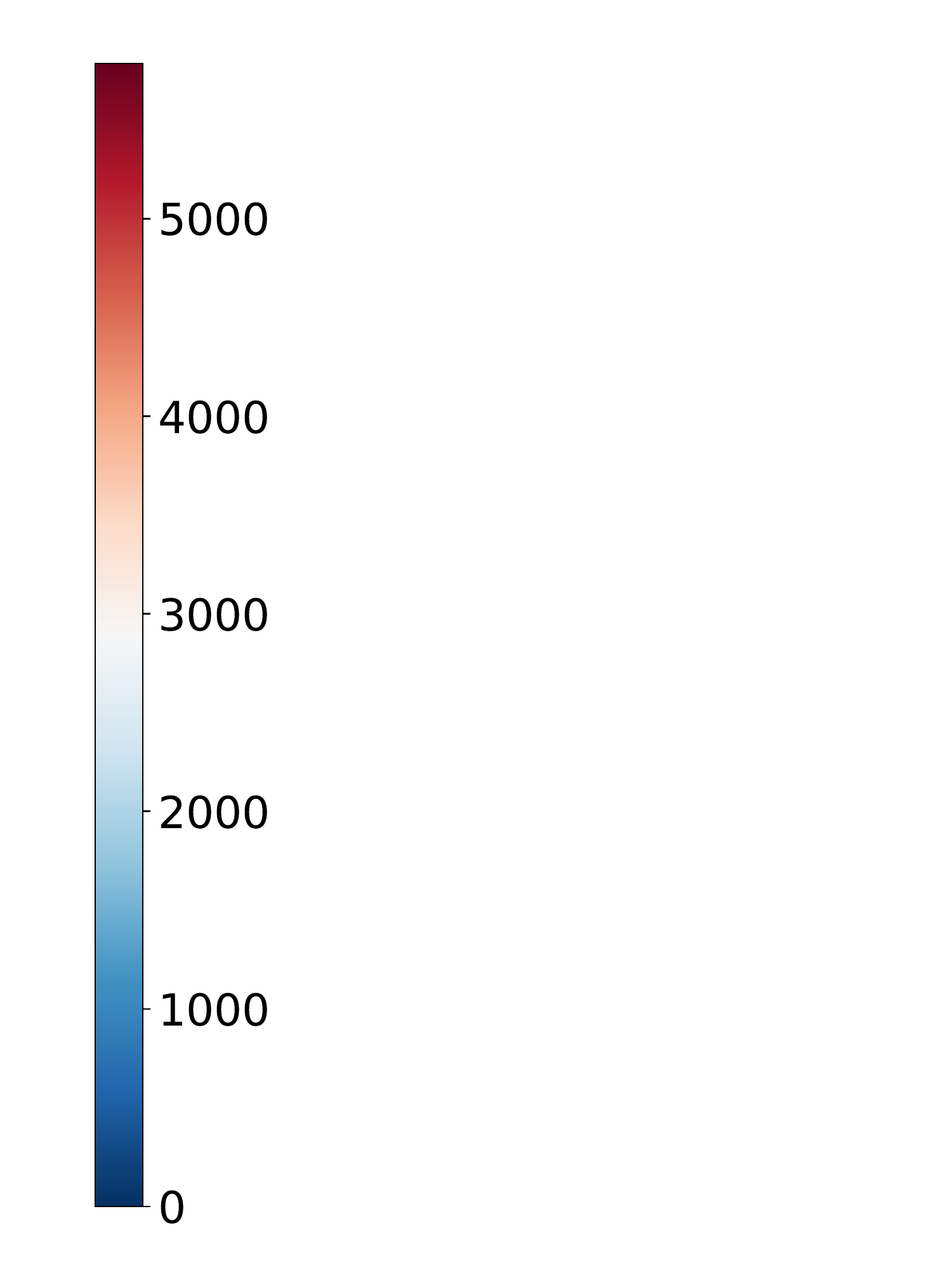}
\subcaption{}
\label{colorbar2}
\end{subfigure}
\caption{Composite telescopic images of all four H.E.S.S. Phase I telescopes for (a) $\gamma$ initiated shower, (b) proton initiated shower at 100 TeV, and (c) the used color scale which is indicative of the photo-electron counts.}
\label{TelImage}
\end{center}
\end{figure}

With the requisite information stored in the detector pixel intensities, the final image can then be stored in intensity color-coded JPEG format. We show some sample images in Fig.~\ref{TelImage} for $\gamma$ and $p$ initiated showers. The four large octagons in each figure correspond to the H.E.S.S. Phase-I telescope cameras. Each telescopic camera image consists of 960 hexagonal detector pixels. The color-coded detector pixel intensities are represented adopting a colorscale shown alongside the figure.

We are using the Python module \texttt{ctapipe},to extract the RGB telescopic images, to perform the calibration of pixel intensities and to apply the image cleaning procedure. Some options that can be set for the final RGB images are, for instance, the image pixel dimensions and the colormap to encode the pixel intensities. We choose a fixed size of $80 \times80$ pixels for the final JPEG image\footnote{These pixels are the pixels of our JPEG image and should not be confused with the detector pixels. However, our choice of number of pixels for the JPEG image is motivated by the number of detector pixels, such that one JPEG image pixel roughly captures information about one detector pixel.}. We have checked that saving in relatively better lossless image formats, or using images with higher resolutions, do not significantly change our figures-of-merit.

The colormap for each telescope is chosen such that the full color range provided by the corresponding library can be used to represent detector pixel intensities in the range from  0 to \texttt{PE$_{\text{max}}$} photoelectrons. For showers in a particular energy range, regardless of the SM primary used to generate them, we use a fixed value of \texttt{PE$_{\text{max}}$}, where the value is chosen to be sufficiently high so that the RGB colors are not saturated by high photo electron counts. For 100 TeV shower images, \texttt{PE$_{\text{max}}$} is set to 5786 photoelectrons (which is the maximum detector pixel intensity in all our simulated shower images).

 Following the procedures delineated above and in the previous subsections, we generate 10000 images each for $\gamma$, $p$, He, and C in the standard set. They will serve as the inputs for our ML algorithms.

\subsection{Anomalous signal events}
\label{sec:anomalysim}
In order to test our anomaly finder, we need some prototypical BSM event images that have subtle differences from the standard images generated by Standard Model primaries.
To this end, we choose to simulate a BSM $\zp$ particle with a mass $m_{\zp} = 1$~TeV which decays in the upper atmosphere to an electron-positron pair. Such $\zp$s are generic in many extensions of the SM~\cite{Rizzo:2006nw}. One could imagine that such a $\zp$ is produced in a cosmic ray collision event, or that it corresponds to some long lived particle that decays in the upper atmosphere. The $\zp$ will be taken to have a large energy (100 TeV), similar to the energies we use for the initial set of standard events. Given the ratio of energy and mass of the $\zp$, this would lead to boosted decay products, so that the resulting $e^-$ and $e^+$ would have a small opening angle $\theta_{\textrm{op}} \sim \frac{ m_{\zp}}{ E_{\zp}} \sim 0.01$~rad between them. The precise value of the $\zp$ mass here is not very important, and has been merely chosen so that it would roughly correspond to a particle near the current limits from the LHC~\cite{ATLAS:2019erb,CMS:2018ipm}.

\begin{center}
\begin{figure}
\begin{subfigure}{0.45\textwidth}
\centering
\includegraphics[height=4cm,width=4cm]{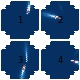}
\subcaption{}
\label{Type-I}
\end{subfigure}
\begin{subfigure}{0.45\textwidth}
\centering
\includegraphics[height=4cm,width=4cm]{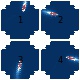}
\subcaption{}
\label{Type-II}
\end{subfigure}
\caption{Z$'$ shower images. The shower image on the left clearly shows two distinct prongs which make the $\zp$ shower image visibly distinct from SM shower images. The shower image on the right is also from a $\zp$, but the two pronged nature is harder to see. The use of image remapping (see Sec.~\ref{sec:remap}) to enhance dim pixels will make the two pronged nature more apparent to the eye, as well as to our autoencoder.}
\label{Zprime_image}
\end{figure}
\end{center}

The behaviour of a shower initiated by such a boosted $\zp$ is similar to that of boosted heavy gauge bosons such as the $W^\pm$ or $Z$ that decay to quarks at the LHC. At the LHC, the jets initiated by these quarks are collimated and can sometimes look like a single ``fat-jet''. Many strategies have been developed to distinguish such fat-jets from regular jets initiated by single quarks/gluons (see for instance~\cite{Ellis:2009su,Thaler:2010tr,Rentala:2014bxa}). Similarly, as we shall see, the images that appear in the telescope for $\zp$ initiated cosmic ray showers could appear visually indistinct from the standard set of images initiated by SM primary particles (see for instance Fig.~\ref{Zprime_image}). Thus our protypical BSM candidate maps to a realistic problem of separating out anomalous events that could visually mimic SM events, but which could be differentiated using advanced ML techniques.

Since a standard $\gamma$ interacting with a nucleus in the upper atmosphere would also typically give rise to an $e^-e^+$ pair, one might wonder why they would lead to different shower patterns. There are, however, important physical differences between the shower of the $\zp$ and that of a $\gamma$, which would lead to differences in the detector images at an IACT. Firstly, the opening angle for the $\zp$ interaction would be wider than that of the $\gamma$ interaction. Second, given plausible model assumptions, the $\zp$ decay to an $e^-e^+$ can occur at various heights in the atmosphere, depending on the $\zp$ production cross-section or decay width. This is in contrast to a gamma shower which would undergo its first interaction approximately 1 radiation length from the top of the atmosphere. Finally, the pair production in the case of a gamma, unlike that of a $\zp$, must occur in the presence of a background nucleus which can also recoil and could also contribute to the observed shower pattern.

The main issue with simulating the shower initiated by such a $\zp$ is the limitations imposed by \texttt{CORSIKA} in the allowed set of primary particles that can be used to initiate a shower --- it allows for $e^-$ and $e^+$ primaries, but not a BSM $\zp$ directly.

\begin{figure}
\centering
\includegraphics[height=10cm, width=14cm]{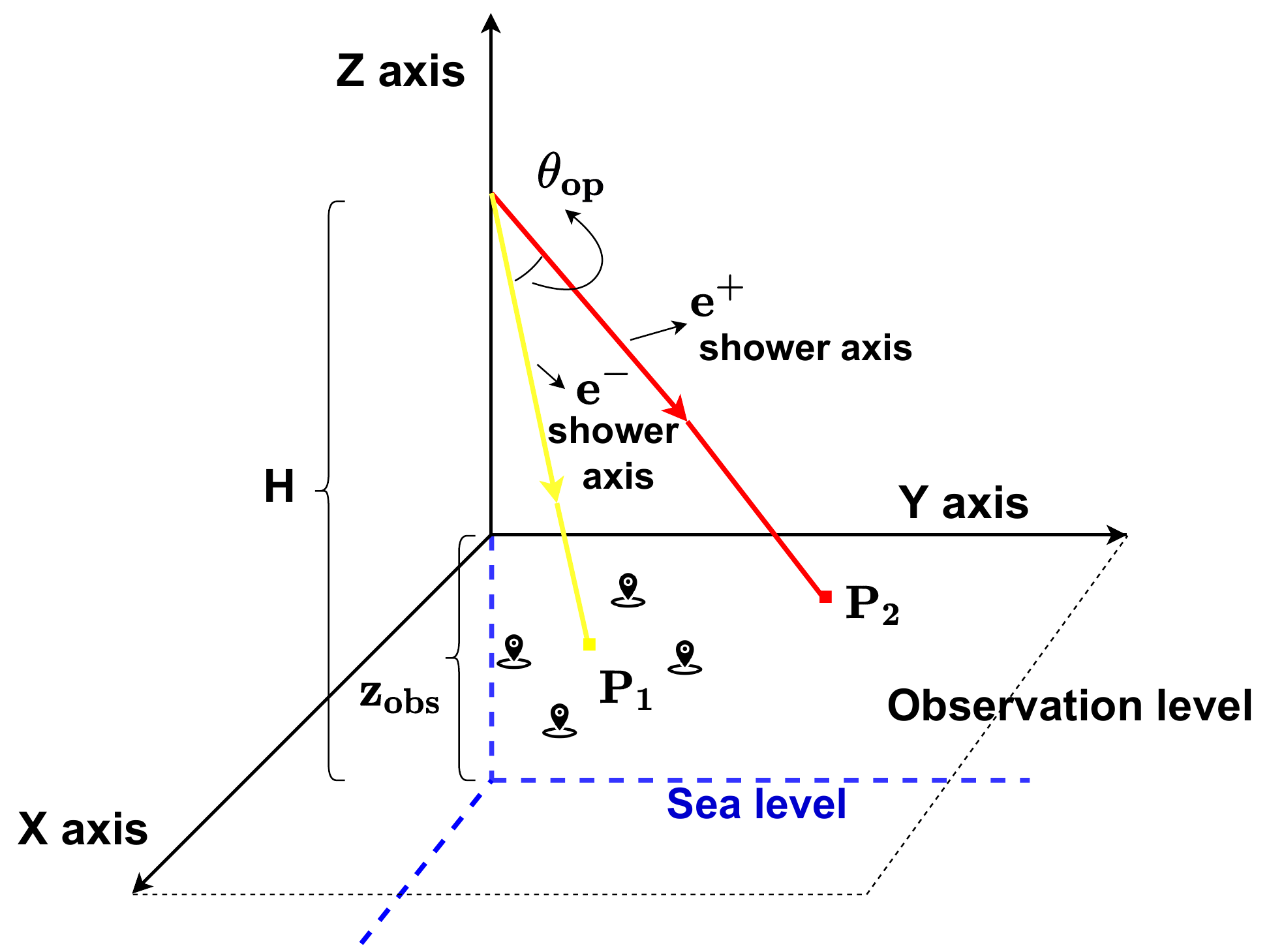}
\caption{The figure shows the physical set up of our simulated $\zp$ events. A high energy $\zp$ decays to an $e^-$ and $e^+$ which have a small opening angle $\theta_\textrm{op}$ between them. The $e^-$ and $e^+$ each initiate a shower, and these two showers need to be separately simulated and the shower images then need to be superposed to obtain the final simulated $\zp$ shower image. The shower axes of these showers intersects the detector plane at the points $P_1$ and $P_2$, respectively. We take the H.E.S.S. array telescopes (marked by geomarkers) to be centered around $P_1$. When simulating the $\zp$ events in \texttt{CORSIKA}, care must be taken to displace the detector center position for the $e^+$, which is by default placed at $P_2$, to the point $P_1$. This ensures that the detectors are at the same physical location for both the $e^-$ and $e^+$ showers.}
\label{Shift}
\end{figure}

To overcome this technical difficulty, we follow the following steps:
\begin{enumerate}
\item The first step in our modified simulation procedure will be to generate $e^-$, $e^+$ pairs from $\zp$ decays in the $\zp$ rest frame, and the corresponding distribution of correlated $e^-$,~$e^+$ momenta . This step is implemented in Madgraph~\cite{Alwall:2014hca} using the $\zp$ in the $B-L$ model (\textit{`B-L-N-4$\_$UFO' file}) \cite{Amrith:2018yfb, Deppisch:2018eth, Basso:2008iv}.

\item The four-vectors of the $\zp$ and its decay products are then boosted such that the $\zp$ has an energy of $E=100$~TeV and makes a zenith angle $\theta_{\textrm{shower}}$ and azimuthal angle $\phi_{\textrm{shower}}$. These angles are chosen in a cone of semi-vertical axis $1.5^\circ$ around the same $\theta_0$ and $\phi_0$ that we choose for SM generated shower images. This procedure mimics the choices of energy and angles made by the standard primaries when using the \texttt{VIEWCONE} option in \texttt{CORSIKA}.

\item For each $\zp$ event, we initialize \texttt{CORSIKA} twice, in a sequential manner. Once with an $e^-$ primary, and then again with an $e^+$ primary. The 4-vectors of the $e^-$ and $e^+$ are correlated and chosen to have an energy and direction corresponding to the result obtained from the step above. In order to ensure that the $e^-e^+$ originate from the same point in the sky, corresponding to the location of the $\zp$ decay, we also set a common height $H$ above sea-level for the starting altitude for both the $e^-$ and $e^+$ propagation. We pick the value of $H$ from a uniform distribution between $5-16$~km\footnote{This range is chosen because it results in the maximum number of secondary particles generated by both the $e^+$ and $e^-$ showers.}. To implement this choice in the \texttt{CORSIKA} input card, we set the value of the grammage corresponding to this height.

\item In our detector simulation we use \texttt{sim\_telarray} and fix the telescope orientations so that the telescopes point in the direction of the cone axis in which the shower lies. We take the shower images obtained in each detector for the $e^-$ and $e^+$ separately, and then superimpose them before performing the cleaning procedure step described previously for the SM initiated shower images. The superposition is performed by adding the photo-electron counts in each of the corresponding detector pixels. This final superposed image after cleaning should correspond to the shower image generated by the $\zp$, as would be seen by the H.E.S.S. telescope.
\end{enumerate}

There is an important and subtle correction which must be taken into account in the last step above. In general in our physical set up, the showers from $e^-$ and $e^+$, originating from the same point in the sky, intersect the observational level at different points. Thus, if say the $e^-$ shower axis intersects the center of the detector system, the $e^+$ shower axis will not. As we described in the previous sub-section, in \texttt{CORSIKA}'s co-ordinate system, the point where the shower axis intersects the observational plane is taken to be $(0,0, \zobs)$. For the standard events, we centered our detector system around this point. For the anomalous $\zp$ events, since we are calling \texttt{CORSIKA} twice for the same $\zp$ event to simulate the $e^-$ and the $e^+$ showers, we must correct the locations of the detectors in the \texttt{CORSIKA} co-ordinate system for at least one of these primaries, in order to ensure that we are simulating a single physical detector system.

For a given event, our convention will be to first choose, with equal probability, either one of the $e^-$ or $e^+$ shower, and to assume that the shower axis for this particle, say $e^-$, intersects the detector plane at the center of the detector system. However, for the other particle $(e^+)$, we will assume that the shower axis intersects the detector plane, off-center from the detector center. This displacement can be seen in Fig.~\ref{Shift}.

In order to ensure that the final JPEG images that we generate for the anomalous set are similar to those of the standard set with SM primaries, we use the same image size of $80\times 80$ pixels, and the same color scale as described in the previous sub-section.

We generate images corresponding to a $100$ TeV $\zp$. Out of the total 15700 showers we simulated, we found only 4000 events which have all four telescopes triggered. Only these 4000 images are selected for our anomalous image bank. In Fig.~\ref{Zprime_image}, we show some of these simulated $\zp$ shower images. The first image, Fig.~\ref{Type-I}, shows a distinct ``two-pronged'' behaviour that visually distinguishes it from the standard images. This two-pronged structure arises because both the $e^-$ and $e^+$ showers are captured simultaneously, but in spatially distinct regions of the detectors. This is similar to boosted event topologies at the LHC. The second image, Fig.~\ref{Type-II}, corresponds to a $\zp$ event that is visually indistinct from a standard image. We discuss an additional image pre-processing step, called remapping which enhances dim features and can thus bring out the two-pronged nature of shower images such as those of Fig.~\ref{Type-II}.

\subsubsection{Image remapping for anomaly finder}
\label{sec:remap}
\begin{figure}
\begin{subfigure}{0.45\textwidth}
\hspace{1.3cm}
\includegraphics[height=4cm, width=4cm]{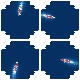}
\subcaption{}
\end{subfigure}
\begin{subfigure}{0.45\textwidth}
\hspace{1.3cm}
\includegraphics[height=4cm, width=4cm]{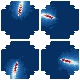}
\subcaption{}
\label{SM_remap}
\end{subfigure}
\hspace{-5cm}
\caption{$\gamma$ shower image (a) before detector pixel intensity remapping and (b) after detector pixel intensity remapping. }
\label{SM_remap_nomap}
\end{figure}

\begin{figure}
\begin{subfigure}{0.45\textwidth}
\hspace{1.3cm}
\includegraphics[height=4cm, width=4cm]{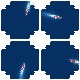}
\subcaption{}
\label{Z_nomap}
\end{subfigure}
\begin{subfigure}{0.45\textwidth}
\hspace{1.3cm}
\includegraphics[height=4cm, width=4cm]{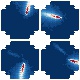}
\subcaption{}
\label{Z_remap}
\end{subfigure}
\caption{$\zp$ shower image (a) before detector pixel intensity remapping and (b) after detector pixel intensity remapping. The two pronged nature of the $\zp$ is clearly visible after remapping. }
\label{Remap_image}
\end{figure}

For testing our autoencoder as an anomaly finder in Sec.~\ref{sec:result_anomaly}, we find it helpful to first remap the detector images for both the standard and anomalous events to enhance the dim (detector) pixels and make them comparable to the brighter pixels. We use a $\sqrt{x}$ type pixel remapping function that was suggested in ref.~\cite{finke2021autoencoders} for the photoelectron counts in each detector pixel. The rescaled detector images are then taken to plot the RGB telescopic images and the colorscale for these remapped images corresponds to detector pixel intensities (in p.e. units) within range from 0 p.e. to $\sqrt{\text{\texttt{PE}}_\text{max}}$.

A specimen of $\gamma$ and $\zp$ shower images before and after remapping are shown in Fig.~\ref{SM_remap_nomap} and Fig.~\ref{Remap_image}, respectively.

From Fig.~\ref{Z_remap}, we see that after remapping of pixel intensities, dim pixels are amplified in comparison to Fig.~\ref{Z_nomap}. The remapped image of $\zp$ shower in Fig.~\ref{Z_remap} clearly shows two-pronged behaviour and is now visually distinct from the remapped SM shower images in Fig.~\ref{SM_remap}. This will make it easier for the autoencoder to identify anomalous events.

We have found that our autoencoder can flag anomalous $\zp$ events with or without image remapping. However, the performance of the autoencoder is slightly better with image remapping.  Hence, we will work only with remapped images when presenting our results for the autoencoder. It is important that we work with both SM images and $\zp$ images which are remapped when training and testing our autoencoder. This is because we will not a~priori know which events are anomalous, and hence all images from the detector have to be remapped in the hope of making anomalous events look more distinct. We choose not to perform this remapping for our binary and multi-category classifiers.

\section{Machine learning architectures and performance metrics}
\label{sec:architecture}

In this section we describe in detail our machine learning architectures. We may pose the role of the various architectures in response to the forms of the three problem statements that we briefly described in the introduction. We clarify the precise problem definitions first.
\begin{enumerate}
    \item \textbf{Binary classification:} Given some typical standard images corresponding to SM particle initiated cosmic ray showers, can we train a machine to predict whether new images fed to it are initiated by a particular SM primary, such as a high energy gamma, or by one of the other standard particles? This would correspond to the typical problem of gamma-hadron separation at an IACT.
    \item \textbf{Multi-category classification:} Given some typical standard images corresponding to SM particle showers, can we train a machine to identify the specific primary that initiated the shower? This would go a step further than simple gamma-hadron discrimination, and would actually be an attempt to identify not just if the event is initiated by a hadron, but also \textit{what type} of hadron is initiating the shower.
    \item \textbf{Anomaly detection:} Given some typical standard images corresponding to SM particles, can we train a machine to flag anomalous events that it has not encountered in training? This would be used as a detection technique to find generic BSM particles such as the $Z^\prime$.
\end{enumerate}

Supervised machine learning techniques are well suited for solving the first two types of problems. For these problems, we feed the machines data with labels --- so that it can learn to identify images of a particular type and classify them as belonging to that type.

Unsupervised machine learning is more suitable for the third type of problem. In this case we train the machines on unlabelled standard images and the machine learns features of the data in such a way that it can flag events that are not similar to those that it has seen in training.

For each problem, we also need to specify the metrics used to judge the performance of the machine towards accomplishing the specific task.

In the previous section, we described the creation of simulated cosmic ray data-sets corresponding to standard and anomalous events. The output from the simulation and image generation phases is represented by a single composite $80\times80$ JPEG image, formed from all the individual H.E.S.S. detectors. As mentioned earlier, we have generated a set of standard images for $\gamma$, $p$, He, and C and a set of anomalous images corresponding to a $Z^\prime$ decaying to $e^-$, $e^+$. These images will be the inputs to our machines.

The ML architectures we utilise have all been implemented using Keras 2.3.1~\cite{chollet2015keras} with Tensorflow 2.2.0~\cite{abadi2016tensorflow} back-end.  For training purpose, we used the ADAptive Moment (ADAM) optimizer~\cite{kingma2017adam} with a batch size of 100 and a mild early stopping criterion with patience $=30$. We have used the \texttt{classification\_report} of \texttt{Sklearn} module in python to evaluate the performance metrics.

In Sec.~\ref{sec:SL}, we discuss the architecture of a convolutional neural network (CNN) that we set up to perform supervised learning for binary and multi-category classification. We also describe some standard metrics to test the performance of these classifiers. In Sec.~\ref{sec:USL}, we discuss the architecture of an autoencoder that we have used for anomaly detection. We describe a different metric that can be used to assess the performance of the autoencoder. We will discuss the actual performance of our machines on our simulated data in Sec.~\ref{sec:Results}.

\subsection{Binary and multi-category classification}
\label{sec:SL}
For our binary and multi-category classification, we use only our standard image sets for both training and testing. The standard set images are passed as labelled data (with labels corresponding to the primary type) to the machine. The ability of the machine to correctly classify these images after learning will quantify the efficiency of the ML architectures to distinguish between conventional CR events.

\subsubsection{Classifier architecture}
\label{sec:SL_architecture}
We use a CNN architecture with a similar structure for both binary and multi-category classification. CNNs are extremely useful for image recognition. Their main advantage is that they preserve the spatial relationship between pixels and they learn the relevant underlying features progressively in each layer of the architecture --- features such as edges, pertinent constituent shapes etc.

\begin{figure}
\centering
\includegraphics[width=15cm, height=5.1cm]{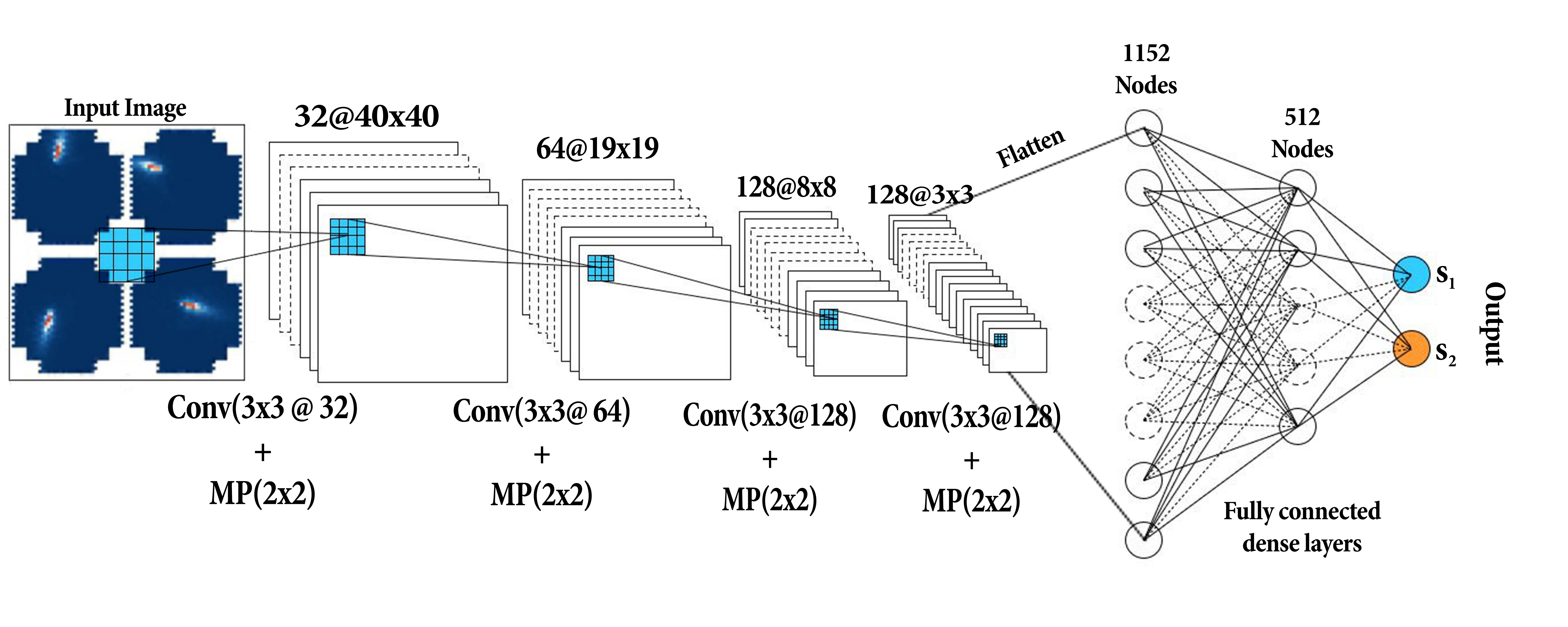}
\caption{Schematic diagram of the CNN architecture used for binary and multi category classification in our work. We use four convolutional layers, denoted as Conv, in the figure. Each convolutional layer has a number of filters with $3\times 3$ convolutional kernels that are used to extract feature maps of the input images. Max pooling (denoted as MP) is applied to reduce the dimensionality of the feature maps at every layer.
For example the input layer takes in an $80\times 80$ RGB cosmic ray shower image and converts it into 32 feature maps of size $40\times 40$ each. The convolutional layers are followed by two of fully connected layers which result in an output layer with nodes labelled by $s_i$. In this figure we represent the output layer for a binary classifier ($i=1,2$ only). For our multi-category classifier, the output layer has as many nodes as the number of categories.}
\label{CNN_arch}
\end{figure}

We now describe the layers of our CNN architecture (see Fig.~\ref{CNN_arch}) that we use for supervised learning below:
\begin{itemize}

\item {\fontfamily{phv}\selectfont Input Layer:} The input to our CNN model is an RGB telescopic image from our ``standard'' set of air showers, of dimension $80\times80$. The RGB color encodes the photoelectron count in the telescopes. We pad the image with 2 extra columns and rows of zeros for each color. Thus our input consists of three $82\times 82$ images, one for each RGB color. The reason for the zero padding will be apparent later.

\item {\fontfamily{phv}\selectfont Convolutional Layers:}
The input is first passed through convolutional layers which are intended to progressively extract the main characteristic features of the input image. Our CNN model has in total four convolutional layers.

Each convolutional layer takes in an input which can be thought of as a collection of $n$, $m\times m$ images. Each $m\times m$ image is called a feature map, and each ``pixel'' of the feature map is a real number. Thus, we can think of the $n$ input images as corresponding to $n$ features. We denote each input image as $I_\alpha$, where $\alpha$ runs from $1$ to $n$. For the first convolutional layer, we have $n=3$ and $m=82$, corresponding to taking in the zero-padded RGB images.

The convolutional layer converts the input feature map into an output feature map of reduced dimensionality. The output of the layer consists of $n^\prime$, $\lfloor \frac{m-2}{2} \rfloor \times \lfloor \frac{m-2}{2} \rfloor$ images. Where $n^\prime$ is the number of features that we extract using this layer. Choosing the values of $n^\prime$ for each layer is part of the definition of the architecture of the CNN. We denote the $\alpha^\prime$-th output image of the layer as $O_{\alpha^\prime}$, where $\alpha^\prime$ runs from $1$ to $n^\prime$. Each output image can be thought of as characterizing the $\alpha^\prime$-th feature of the input image.

The conversion from input images to an output image in a given convolutional layer proceeds through 4 steps,
\begin{enumerate}
    \item convolution,
    \item application of bias,
    \item application of an activation function,
    \item application of a max pooling layer.
\end{enumerate}

Our convolutional layers have $n\times n^\prime$ \, $3 \times 3$  convolutional kernels and $n^\prime$ bias parameters. We denote each convolutional kernel as $M_{ \alpha^\prime\alpha,}$, where $\alpha$ ($\alpha^\prime$) runs from $1$ to $n$ ($1$ to $n^\prime$), and we have suppressed the explicit $3\times 3$ indices of each kernel. The bias parameters are denoted as $b_{\alpha^\prime}$. Thus, in total the layer has $(3\times 3)(n\times n^\prime) + n^\prime$ parameters. These parameters are ``learnable'' in the sense that the machine will iterate over these in training in order to find some optimal parameters for classification of input images.

Symbolically, after convolution and application of bias, the $\alpha^\prime$-th output map is related to the input maps viz. \mbox{$O_{\alpha^\prime} = \sum_\alpha M_{\alpha^\prime\alpha} * I_\alpha  + b_{\alpha^\prime}$}, where $*$ denotes convolution. At this stage, the output image is $m-2 \times m-2$ dimensional. The combined operations of convolution and application of bias are referred to as application of a filter. Thus, there are as many filters as output feature maps in a given convolutional layer.

To this output image we now apply a \texttt{Rectified Linear Unit (RELU)}  activation function. Where the RELU function is given by,
\begin{equation}
RELU(x)=
\begin{cases}
x \qquad \textrm{for}~~ x>0,\\
0 \qquad \textrm{for}~~ x<0.
\end{cases}
\end{equation}

Finally, we apply a max pooling layer to reduce the dimensionality of the output image. The max pooling layer simply coarse grains each output image by taking the maximum of distinct $2\times2$ blocks of each output. This reduces the image size to $\lfloor \frac{m-2}{2} \rfloor \times \lfloor \frac{m-2}{2} \rfloor$.

For our four convolutional layers, the layers have value of $n^\prime =32, 64,128,128$, respectively. The first convolutional layer takes the original (padded) cosmic ray image\footnote{We can now understand why the original cosmic ray image needs to be padded with 2 additional rows and columns. Since the $3\times3$ kernels are convolved with the input image, we would like a unique position for convolving the kernel for every pixel of the unpadded original input.} as input and the output of this layer is passed to the next as input and so on. As the image is passed from one convolutional layer to the next, we generate more feature maps of smaller image size that should contain only the essential features of the original image.

\item {\fontfamily{phv}\selectfont Flatten and Fully Connected (FC) Layer:} At the end of the fourth convolutional layer, we flatten all the images and get a single 1D array with 1152 nodes. This flattened layer is then fully connected to a dense layer with 512 nodes (see Fig.~\ref{CNN_arch}),

Our fully connected layer takes in $n$ inputs $x_i$ ($i=1..n$) and gives $n^\prime$ outputs $y_j = \sum_i m_{ji} x_i + b_j$, where $j=1..n^\prime$. The coefficients $m_{ij}$ and $b_j$ are machine parameters to be learnt. Thus for a fully connected layer there are $n\times n^\prime + n^\prime$ parameters. We also apply the RELU activation function to our fully connected layers, except for the last layer for which we use the softmax activation function (see below).

\item {\fontfamily{phv}\selectfont Output Layer:} Finally we fully connect the dense layer to our output layer. The output layer has $N_{\textrm{category}}$ nodes, where $N_{\textrm{category}}$ is the number of labelled categories (for e.g. for the binary classifier $N_{\textrm{category}}=2$, whereas for multi-category classification of the standard particles $\gamma$, $p$, He, C, we choose $N_{\textrm{category}}=4$). For the last FC layer that connects to the output layer, we use the softmax activation function. This activation function takes in a vector of inputs $y_i$, where $i = 1..N_{\textrm{category}}$ and returns a vector $s_i$ where,
\begin{equation}
    s_i = \frac{\textrm{exp}(y_i)}{\sum\limits_{i=1}^{N_{\textrm{category}}} \textrm{exp}(y_i)}
\end{equation}
We store $s_i$ as the output of the machine in the $i$-th output node.
\end{itemize}

When we train our machine on labelled input data, the images are passed as data to the machine at the input layer, as described above. The labels are passed as vectors of dimension $N_{\textrm{category}}$ to the machine using the ``one-hot encoding'' method---for example for $N_{\textrm{category}}=4$, images belonging to category 1 (2) are labelled with a vector $L = (1,0,0,0)$ $(L = (0,1,0,0))$ and so on. These labels will be used by the machine to calculate a loss function.

For our architecture, we will choose the \texttt{CategoricalCrossentropy} loss function. This loss function is defined as,
\begin{equation}
\text{Loss}=-\sum_{i=1}^{N_\textrm{category}} L_i \cdot \textrm{log } s_i,
\end{equation}
where $L_i$ is the $i$-th value of the label vector. By virtue of the softmax activation function, the output vector $s_i$ has positive entries, with sum normalized to unity. These values $s_i$ can be interpreted as the probability that a given image is of a particular type labelled by $i$. Thus the loss function has the interpretation of a relative entropy (or likelihood) between the true labels and the reconstructed (probabilistic) labels. During training over all the input categories, the machine optimizes the variable parameters (such as convolution kernel parameters, bias parameters), in order to minimize the loss function averaged over all training inputs.

Once the machine has been trained we can validate  the performance on a validation data set to check that the variable parameters have converged and the performance is stable, i.e. the amount of information learned about the images is nearly saturated.  Once this is done, we are finally ready to test our machine performance on test data to assess the machine's performance for the task of classification.

For our validation and testing phases, we pass an unlabelled test image (for which we know the correct category) to the machine and check the output vector $s_i$. We take the classification made by the machine to be the category corresponding to the label $i$ for which $s_i$ is maximum. We can then check how often the machine correctly classifies the test input images. The performance on this testing data set is used to quantify the performance of the ML architecture.

We will describe in detail the training, validation, and testing of our machine's performance on the applicable datasets in Sec.~\ref{sec:Results}. In preparation for this, it will be useful to describe here some metrics to evaluate the classifier's performance during the testing phase. We do this next.

\subsubsection{Metrics to evaluate the performance of our classifier}
\label{sec:class_metrics}
A classification metric is a number that helps us assess the performance of a trained classifier model, on a testing dataset --- one that it has not seen during the training phase. A variety of classification metrics are used in the machine learning literature. Here, we describe briefly the ones that we will use when presenting our results in Sec.\,\ref{sec:Results}.

\begin{itemize}
\item \textbf{Binary classification metrics:} For evaluating our binary classifier, we use the ``Accuracy Score'' as the classification metric. Accuracy is the ratio of number of correctly classified instances to the number of total instances on which the classifier is tested, i.e.
\begin{equation}
\text{Accuracy}=\frac{\text{Number of correctly classified instances}}{\text{Total number of instances}}.
\label{eq:accuracy}
\end{equation}
Here an instance describes a particular test image, which belongs to one of either of the two categories on which the binary classifier had been trained.

\item \textbf{Multi-category classification metrics:} In the case of multi-category classification, accuracy score is not an appropriate metric for evaluating the performance of the classifier. For instance, suppose in our testing set there are `$n_a$' number of images of type~A, `$n_b$' number of images of type~B, and `$n_c$' number of images of type~C. Now, the classifier has correctly identified the category of say `$m$' of the total number of images. The accuracy score would then be $\frac{m}{n_a+n_b+n_c}$. However, this score does not give us full information about the performance of the machine. It could have been the case that nearly all A type images are classified well by the classifier model, whereas the classification of images of type B is very poor, and perhaps that of type C is mediocre. Thus, in the multi-category case, better metrics to quantify the efficacy of the ML classifier are called for.

A more complete quantification of the performance is given by the so-called confusion matrix table. An example of the confusion matrix is shown in Table \ref{Conf_mat}, for the case of 3 categories.
\begin{table}
\centering
\begin{tabular}{|m{1.5cm}| m{0.5cm} m{2cm} m{2cm} m{2cm}|}
\hline
 & \multicolumn{4}{c|}{Predicted values}\\
  \multirow{4}{3em}{ Actual values}& & A & B & C\\ \hline
 & A & {\fontfamily{qcr}\selectfont TA} & {\fontfamily{qcr}\selectfont FB(A)} & {\fontfamily{qcr}\selectfont FC(A)}\\
 & B & {\fontfamily{qcr}\selectfont FA(B)} & {\fontfamily{qcr}\selectfont TB} & {\fontfamily{qcr}\selectfont FC(B)}\\
 & C & {\fontfamily{qcr}\selectfont FA(C)} & {\fontfamily{qcr}\selectfont FB(C)} & {\fontfamily{qcr}\selectfont TC}\\
 \hline
\end{tabular}
\caption{The confusion matrix for multi-category classification. {\fontfamily{qcr}\selectfont TA} is number of A type image that are correctly classified as belonging to A type. {\fontfamily{qcr}\selectfont FA(B)} is the number of B type images falsely classified as A type images. Other entries of this table are similarly defined.}
\label{Conf_mat}
\end{table}

The rows of the confusion matrix correspond the the actual categories of the image, and the columns correspond to the category predicted by the machine. The entries of the matrix tell us the number of images of the true category which are classified as belonging to a predicted category. For example in this table, {\fontfamily{qcr}\selectfont TA} gives the number of A type images that are correctly classified as A type. {\fontfamily{qcr}\selectfont FA(B)} and {\fontfamily{qcr}\selectfont FA(C)} give the numbers of B type and C type images, respectively, that are wrongly tagged as A type images by the classifier model.
A similar convention is followed for the other terms in the confusion matrix.

Note that the total number of A type images is $n_a= TA + FB(A) + FC(A)$, and so on for types B and C. One disadvantage is that sometimes the confusion matrix per se is hard to interpret directly in terms of machine learning performance. This is because of its reliance on absolute numbers which would in turn depend on the number of instances of events of each category in the testing set.

To mitigate some of the disadvantages of the confusion matrix, more intuitive metrics can be found that represent the performance of the machine learning architecture, in terms of relative numbers. We use the metrics accuracy, precision, recall, and f1-score to characterize our machine performance. We define these metrics below and we will try to give some intuition for what aspect of the performance they indicate. These other metrics can be defined in terms of entries of the confusion matrix, and we will present their definitions for a 3 category classifier in terms of the entries of the $3\times3$ confusion matrix above. The generalization to higher number of categories is straightforward.

The definition of accuracy is similar to that of the binary classifier. It is the ratio of the number of instances correctly classified to the total number of instances. So, from the confusion matrix we have
\begin{equation}
\text{Accuracy}=\frac{\text{\fontfamily{qcr}\selectfont TA}+\text{\fontfamily{qcr}\selectfont TB}+\text{\fontfamily{qcr}\selectfont TC}}{\text{Total number of A, B, \& C type images}}.
\label{eq:acc}
\end{equation}

Precision and recall are metrics which are defined for a particular category (say A). For that category, precision is defined as the ratio of correctly identified images in category A divided by the total number of images (either correctly or incorrectly) classified as belonging to category A. Recall is defined as the ratio of correctly identified images in category A divided by the total number of images in category A.
Thus, precision is a measure of how well we can trust the output of the machine when it tells us that an event belongs to category A. Recall is a measure of how often the machine will correctly classify inputs belonging to category A.

In terms of the entries of the confusion matrix they are defined as,

\begin{equation}
 \text{Precision}~|_{\text{for A}}= \frac{\text{\fontfamily{qcr}\selectfont TA}}{\text{\fontfamily{qcr}\selectfont TA}+\text{\fontfamily{qcr}\selectfont FA(B)}+\text{\fontfamily{qcr}\selectfont FA(C)}},
 \label{eq:prec}
\end{equation}
\begin{equation}
\text{Recall}~|_{\text{for A}}= \frac{\text{\fontfamily{qcr}\selectfont TA}}{\text{\fontfamily{qcr}\selectfont TA}+\text{\fontfamily{qcr}\selectfont FB(A)}+\text{\fontfamily{qcr}\selectfont FC(A)}}.
\label{eq:rec}
\end{equation}

Another metric that is used is the f1-score, which is the harmonic mean of the precision and recall,
\begin{equation}
 \text{f1-score}~|_{\text{for A}}= 2~\frac{\text{Precision}~\times~\text{Recall}}{\text{Precision}+\text{Recall}}.
 \label{eq:f1}
\end{equation}

To reduce the proliferation of performance metrics which are defined individually for each category, one can also define the weighted average of these scores. For example the precision weighted-average is defined by weighting the precision score for each category by the number of images in that category. For example for our 3-category classifier,
\begin{equation}
    \textrm{Precision (weighted average)} = \frac{\sum\limits_{i = A, B, C} \left(n_i \times  \textrm{Precision} |_{\textrm{for i}} \right )}{n_A + n_B + n_C}
\end{equation}
A similar weighted average can be done for recall or f1 score.
\end{itemize}


\subsection{Anomaly detector}
\label{sec:USL}

\textit{Anomaly detection} is a method to identify unusual patterns that do not conform to expected behaviour. Unlike supervised learning algorithms which work with labelled data sets, we use unsupervised learning techniques for anomaly detection. This is motivated based on the philosophy that we do not know a priori what form new physics might take. Thus, it is prudent to develop model agnostic strategies to look for such exotic events in cosmic ray showers. In many cases these exotic events may mimic conventional cosmic ray air showers, and it may not be easy to identify them as anomalous just by a visual inspection.

We make use of autoencoding~\cite{10.5555/3153997}, which is an unsupervised ML technique that first efficiently compresses input data into a lower dimensional parameter space and subsequently attempts to reconstruct the original image as closely as possible from the compressed version. If the resulting image resembles the original input within some tolerance, the image is classified as ``normal'', otherwise the image is classified as ``anomalous''. This paradigm therefore forces the autoencoder to learn the relevant features of a set of training images very well. Crudely speaking, if the standard SM induced showers are taken as the training set, and the autoencoder learns relevant features of these images, then it will be well poised to identify general complements of this set, i.e. general BSM induced cosmic ray showers, whose exact frameworks and mechanisms may as yet be unknown to us.

The difference between this technique and the classifier, is that in this case the machine is only trained on SM images and has not seen any anomalous type of images, before the testing stage. This is why this technique falls under the category of unsupervised learning; the machine learns what can be classified as normal, and thus can flag events which are anomalous -- those with patterns that do not correspond to the patterns that it has learnt while training.

\begin{figure}[H]
\centering
\includegraphics[scale=0.09]{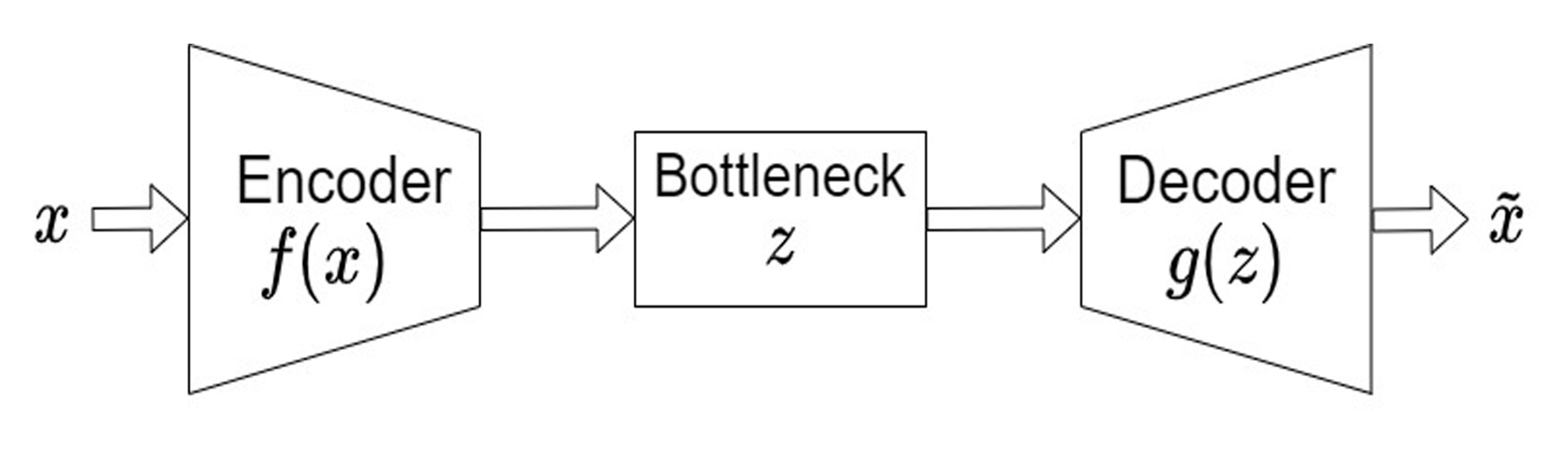}
\caption{Basic structure of an autoencoder.}
\label{Auto_Enc1}
\end{figure}

The basic structure of an autoencoder is shown in Fig.~\ref{Auto_Enc1}. The autoencoder consists of three parts,
\begin{itemize}
\item {\fontfamily{phv}\selectfont Encoder} - This block compresses the input data (denoted as $x$) into a lower dimensional representation (called a latent representation and denoted as $z$) and thereby encodes it, i.e. $z = f(x)$.
\item {\fontfamily{phv}\selectfont Bottleneck layer} - This layer contains the compressed representation ($z$) of the input data. Since the representation is of much smaller size than the input data, this layer tries to encode only the most relevant and important features of the input data.
\item {\fontfamily{phv}\selectfont Decoder} - This block attempts to reconstruct the original data from the lower-dimensional encoded representation. We denote the reconstructed image as~$\tilde{x} = g(z)$, where $g$ is the functional representation of the decoder.
\end{itemize}
The goal of the autoencoder is to construct an output image $\tilde{x}$ that closely resembles the original input image $x$, by using only a compressed representation $z$ of the input. The difference between the original and reconstructed standard images will be the quantity to be optimised over during training.

In order to quantify how similar or dissimilar the output is from the input,  we define a suitable loss function, $\mathcal{L}(x, \tilde{x})$. We use the $\textit{mean squared error (MSE)}$ loss function given by,
\begin{equation}
\mathcal{L}(x, \tilde{x}) = MSE = \frac{1}{m}\sum_{i=1}^{i=m}(x_i - \tilde{x}_i)^2\;.
\label{mse}
\end{equation}
Here, the sum $i$ runs over the corresponding pixels (in all three color features RGB) of the input and output images.

During the training phase, the autoencoder is fed images from the standard set, and it attempts to adjust some learnable machine parameters in order to minimize the loss function. Once the training phase is over, the trained machine can now act as a potential anomaly finder. If it is now fed an anomalous image, it will attempt to compress and encode the image in the same way as it had learnt to compress and then reconstruct standard images. However, for sufficiently different anomalous images, this compression will obviously not be able to capture all the features of the anomalous image. Thus, after reconstruction from this lossy compression, the output of a good autoencoder should yield a high reconstruction error for the anomalous image. However, if the autoencoder is fed standard images, similar to the images that it has been trained on, the autoencoder should yield low reconstruction errors.

In our work, we consider shower images initiated by gamma, proton, helium and carbon as the prototypical standard events (or background events) and images coming from the Z$^\prime$ initiated shower as the prototypical anomalous events (or signal events).  We design an autoencoder that will act as an anomaly finder for cosmic ray events initiated by non-standard or anomalous events. In the next sub-sections, we describe the architecture of the autoencoder, and then we describe some metrics to judge its performance.

\subsubsection{Autoencoder architecture}
\label{sec:ae_arch}
\begin{figure}
\centering

\includegraphics[width = 15.5cm, height = 4.5cm]{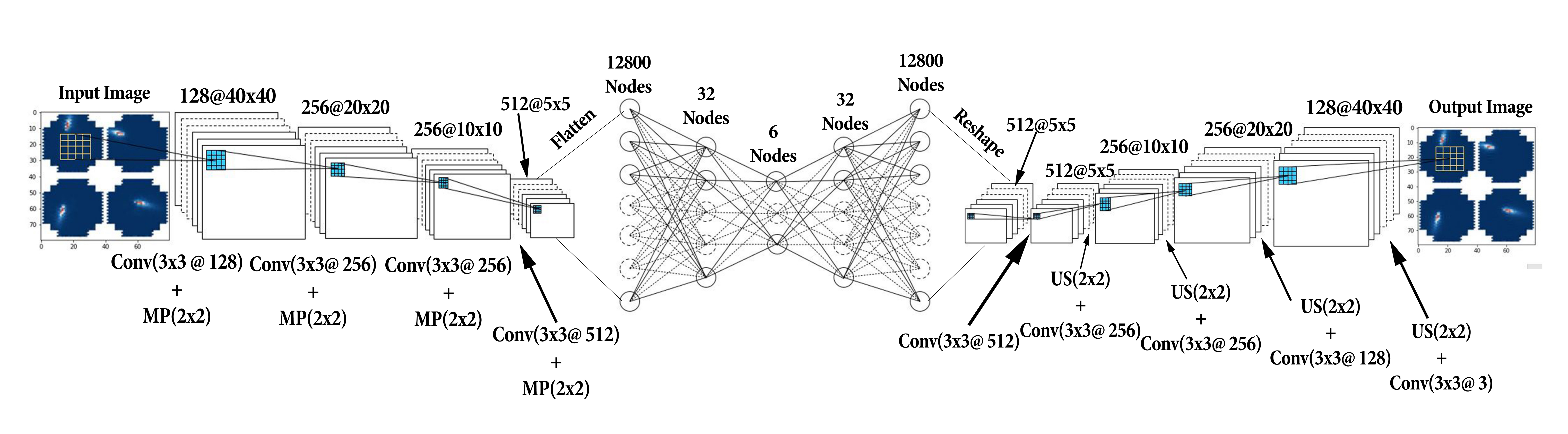}

\caption{The schematic diagram of our autoencoder's architecture. The encoder part consists of 4 convolutional layers and 2 fully connected layers. The convolutional layers are denoted by Conv in the figure. The first convolutional layer has 128 filters, the second and third convolutional layers have 256 filters each, and the fourth one has 512 filters. All the filters have convolution kernels of size $3\times3$. Max pooling (denoted as MP) is applied at each layer to reduce the dimensionality of the feature maps from each convolutional layer. The convolution layers are followed by two fully connected layers with 12800 nodes and 32 nodes, respectively. The bottleneck layer is a FC layer with only 6 nodes. The decoder part reverses the behavior of the enconder, with 2 FC layers followed by a single convolutional layer, which is further followed by four convolutional layers with up-sampling (denoted as US). Up-sampling is used to increase the dimensionality of the decoded feature maps. The final layer gives the decoded output image which can be compared to the original input image to check for reconstruction losses.
}
\label{AE_Arch}
\end{figure}

The schematic diagram of the autoencoder architecture we are using in our work is shown in Fig.~\ref{AE_Arch}. The architecture we employ is a modification of the VGG16 architecture \cite{simonyan2015deep}.

The encoder we implement consists of 4 convolutional layers with down-sampling (reducing image size), followed by 2 fully connected layers which then lead up to a bottleneck layer with 6 nodes. This bottleneck layer with 6 nodes in the middle of the architecture is designed to encode the compressed information pertaining to the input image. The decoder reverses the behaviour of the encoder, with 2 fully connected layers, followed by a single convolutional layer, which is further followed by 4 convolutional layers with up-sampling (increasing image size) implemented before convolution, leading then to the final reconstructed output image. We describe in more detail the design of each layer below.

The input to the autoencoder is once again an $n= 3$ (RGB color), $m \times m = 80 \times 80$ (pixels per color) image. This image is passed to the encoder to be encoded into the bottleneck layer before being decoded. The first part of the encoder architecture consists of a set of four convolutional layers. Similar to the convolutional layers used in our classifier architecture, each convolutional layer uses a set of $n\times n^\prime$ $3\times3$ convolutional kernels, with $n^\prime$ bias terms, and an activation function. Here, as before, our notation assumes that there are $n$ input, and $n^\prime$ output feature maps (which are passed as inputs to the next convolutional layer). We also use $2\times2$ max pooling to reduce the image size from one layer to the next. Our four convolutional layers output  $n^\prime = 128,256,256,512$ feature maps, respectively. These feature maps capture more and more subtle features of the original input image.

There is a slight difference between the convolutional layers that we use here for the encoder and that of our classifier that we described earlier in this section. In the encoder, we zero pad the output of each layer with two extra columns and two extra rows. This ensures  that for each layer, the output images are reduced in size by exactly a factor of 2 along each input image direction. Thus, if a layer takes in $n$ images of size $m\times m$, it outputs $n^\prime$ images of size $m/2\times m/2$ ($m$ is always even for our architecture). All convolutional layers use the RELU activation function.

The output of the last convolutional layer is flattened into a layer with 12800 nodes. This is then fully connected to a 32 node layer, which in turn is fully connected to the bottleneck layer with 6 nodes\footnote{We have also tried to work with a bottleneck layer with 4 or 8 nodes, but we found optimal performance for anomaly detection with the 6 node architecture.}. The fully connected layers all also use the RELU activation function.

Once the image is encoded in the bottleneck layer, the rest of the architecture is designed to decode this information and reform the original image as  accurately as possible. Our decoder reverses the behavior of the encoder. We first have two fully connected layers of 32 and 12800 nodes. The output of the 12800 node is reshaped into 512 square images of dimension $5\times5$, essentially implementing in reverse the flattening step of the encoder. In order to obtain 512 $5\times5$ images that have the interpretation of being feature maps, we follow the reshaping step by a single convolutional layer with a RELU activation function that takes these square images and converts them into 512 $5\times5$ feature maps.

This is then further passed through a series of 4 convolutional layers with up-sampling. The key difference going from the encoder to the decoder is that in these convolutional layers, rather than down-sampling the image by using max pooling, we need to up-sample the images to increase the image size back to a reconstructed $80\times80$ RGB image. The up-sampling that we do, simply takes a pixel after convolution, and replaces it by a $2\times 2$ grid of pixels each containing the same original pixel value. This up-sampling is performed before the convolution step. Thus, the convolutional layers of the decoder takes an input with $n$ images of $m\times m$ pixels and outputs $n^\prime$ images of $2m\times 2m$ pixels. Since we have reversed the order of the convolutional layers we have $n^\prime = 256, 256, 128, 3$ for each of these four convolutional layers. As with the encoder, our decoder also uses $3\times3$ convolutional kernels with a bias and an activation function. Every layer, except the last one, uses the RELU activation function. The last layer, which is connected to the output, uses the sigmoid activation function which is given by,
\begin{equation}
    \Phi(z) = \frac{1}{1+\textrm{exp}(-z)}.
\end{equation}
The final output of the decoder results in an $80\times 80$ RGB output image that can be compared to the input image\footnote{The sigmoid activation function results in an output between 0 and 1 for each RGB channel. In order for the output to be similar to the input, we scale the input image color values to also lie between 0 and 1 before feeding them to the autoencoder. However, when presenting detector images in our results section, the input and output images are rescaled to the standard color scale with values between 0-255 in each color channel.}.

Note that in contrast to the original VGG16 architecture, our architecture uses fewer convolutional layers in both the encoder and the decoder parts to extract specific features from the images. Typically, for a deeper neural network with more layers, we should get better performance. However, deep neural networks comes with the cost of higher computational power. While choosing the number of convolutional layers, we have tried to seek a balance between optimizing performance and avoiding prohibitively large computational costs.

During training, the autoencoder is fed only images from the standard set, and the convolution coefficients and bias parameters are progressively learnt, in such a way as to minimize the loss function or the mean-squared error between the output and input images.

\subsubsection{Figure of merit for autoencoder}
\label{mse_ROC_AUC}
After the autoencoder is trained on standard data, we then feed it test data consisting of both standard and anomalous images. For each image, the autoencoder attempts to reconstruct an output image and then compares it with the input image. The machine then computes a mean-squared error (MSE) difference between the input and output images, as defined in Eq.~\ref{mse}. Standard images should have a low reconstruction error and are thus expected to have low MSEs, whereas anomalous events are expected to have higher reconstruction errors.

We need to define a threshold MSE (which is arbitrary), which we denote as MSE$_\textrm{th}$, that will help us tag an event as anomalous. If the MSE for a particular image is smaller than MSE$_\textrm{th}$, we classify the image as standard type, and if it exceeds MSE$_\textrm{th}$, we classify it as anomalous.

Since we are interested in flagging anomalous events as our signals, we can define two performance quantification metrics, the true positive rate (TPR) and the false positive rate (FPR). For a given MSE$_\textrm{th}$, these are defined as,
\begin{align}
\text{\texttt{TPR}} &=
\textrm{Fraction of BSM shower images correctly tagged as anomalous}, \nonumber \\
\text{\texttt{FPR}} &=
\textrm{Fraction of standard shower images incorrectly tagged as anomalous}. \nonumber
\end{align}

We can plot a Receiver Operating Characteristic (ROC) curve between the TPR and FPR as we vary the threshold MSE$_\textrm{th}$ for anomaly detection. The choice of MSE$_\textrm{th}$ that is to be applied in a particular experimental analysis will depend on the rate of expected anomalous events and the error tolerance for flagging normal events as anomalous. The ROC curve can help the experimentalist pick out the choice of threshold needed for their analysis. A figure-of-merit that can be used to judge the performance of the autoencoder is the ``area under the ROC Curve'' (AUC)  which gives a measure of separability between anomalous and non-anomalous images. The AUC takes values between 0~and~1, and the higher the value of the AUC, the better is the performance of a trained autoencoder model.

\section{Results}
\label{sec:Results}

In the previous sections, we described our simulation of cosmic ray showers and generation of telescopic images seen at H.E.S.S. for standard cosmic ray showers initiated by SM primaries ($\gamma$, $p$ , He, C), as well as for anomalous showers initiated by $Z^\prime \rightarrow e^+e^-$. We also described our machine learning architectures which can be trained to discriminate between the various image types. The problems that we are trying to address are of two types a)~supervised learning - involving learning of labelled image types from training data with the goal of being able to categorize new test images which belong to one of the training categories b)~unsupervised learning - involving learning features of unlabelled training data with the goal of being able to flag anomalous events in test images which are dissimilar from the training data. We also described figures-of-merit for each set of problems that can be used to quantify the performance of our machines.

In this section, we will describe the process of training our machines for the specific tasks, and then we will show their performance results on test data. In the first subsection below, we will discuss the case of supervised learning, specifically our binary and multi-category classification schemes. Then, in the second subsection, we will showcase the results of our anomaly detection method.

\subsection{Supervised learning and classification problems}
\label{sec:result_classification}

In this section, we will discuss the training and performance of our binary and multi-cateogry classifier. For the discussion that follows, we will discuss the training and performance metrics of our binary and multi-category classifier using images where the SM primaries have an energy centered around $E = 100$~TeV with zenith and azimuthal angles selected in a $1.5^\circ$ cone around $\theta_0=0^\circ$ and $\phi_0=0^\circ$ (see Sec.~\ref{sec:EAS_sim}). Results for other choices of energies and angles are presented in appendix~\ref{appendix_classification}.

\subsubsection{Binary classification performance}
\label{sec:binary_result}

The goal for the binary classifier is to identify the categories of testing data which belong to one of two classes.

We first select cosmic ray showers images from any two categories of images in our standard image set, for e.g. gamma and proton images. We have 10000 images for each of the two categories of SM shower images. This set is then split into 8100 training images, 900 validation images, and 1000 testing images for each of the SM primaries. The testing images are not seen by the classifier at any point during the training and so the performance metrics using these test images gives an accurate reflection of the classifier's capability for distinguishing between images initiated by different primaries.

Our binary classifier is trained on the $(8100+8100)$ labelled cosmic ray showers images initiated by two different types of SM primaries. We use the mini-batch gradient descent method to optimize our machine parameters during training. This method of training of the classifier is an iterative process. First, the entire training data set is randomly split into batches of 100 images. After each batch is processed by the classifier, the machine calculates the total loss for the batch and then updates the parameters using the gradient descent procedure. For a sufficiently small batch size, the noise in the loss function can be sufficient to ensure that the machine parameters are not trapped in a local minimum. The training is continued until all training images have been encountered at least once by the classifier. This entire process is referred to as one epoch. Once we have completed an epoch, we can compute an accuracy score (see Eq.~\ref{eq:accuracy}) for the entire training data set, and also for the validation data set which has 900 + 900 images.

The training set is once again randomly split into batches of 100 images and the training is performed again with the new parameters from the previous epoch as seed values for a new training epoch. Once again, we can compute the training and validation accuracy scores at the end of this epoch.

This process is continued until the accuracy score for the validation set does not exceed the accuracy score of an epoch number $i_\textrm{crit}$ for 30 more consecutive epochs. This is known as the `early stop' criterion. Stopping the training at this stage ensures that we avoid over-training. We then take the final machine parameters to be those of the epoch $i_\textrm{crit}$ which has the largest validation set accuracy score.

\begin{figure}
\begin{subfigure}{.5\textwidth}
  \centering
  \includegraphics[width=1.1\linewidth , height=5cm]{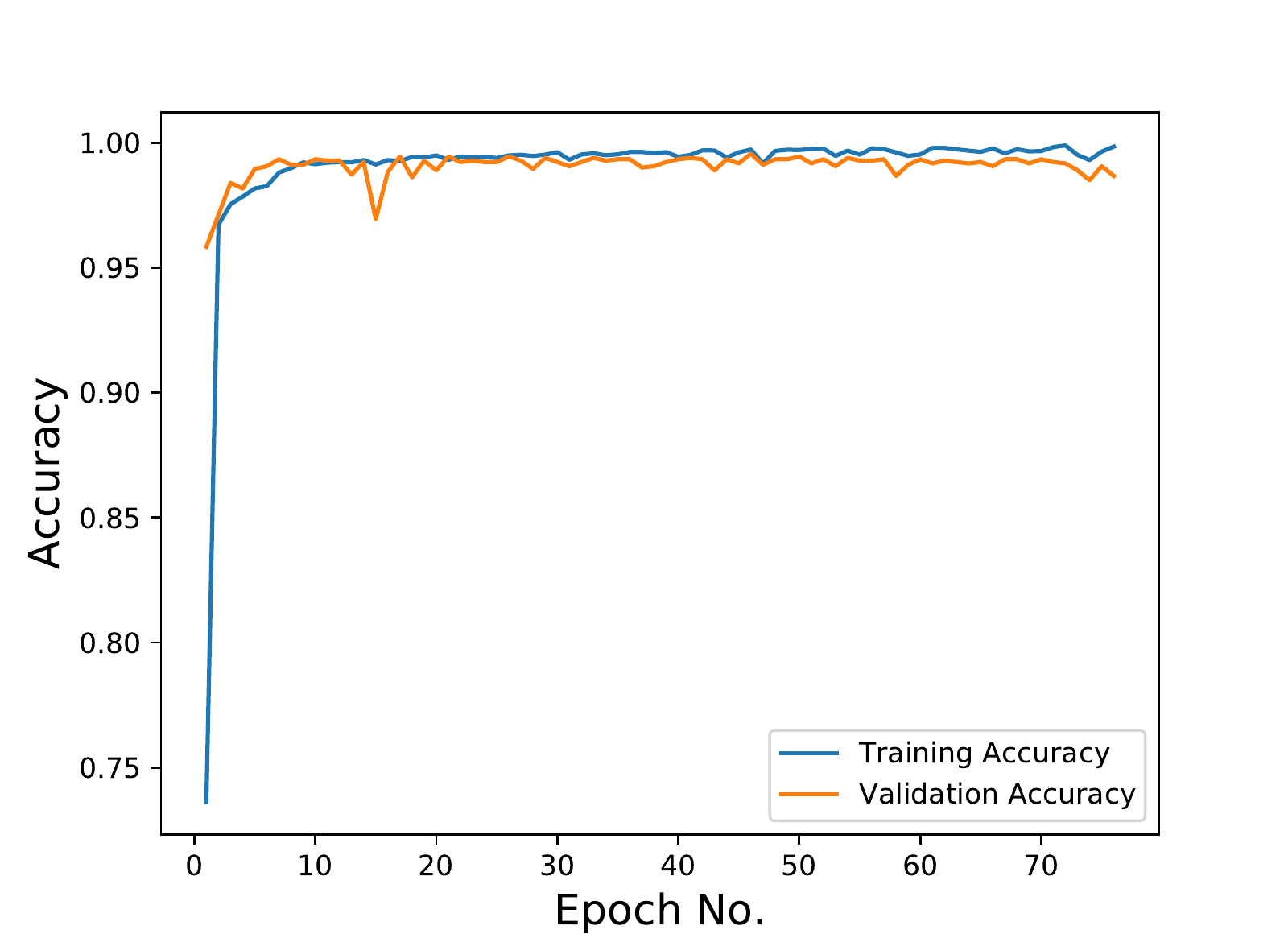} \quad
  \caption{}
\end{subfigure}
\begin{subfigure}{.5\textwidth}
  \centering
  \includegraphics[width=1.1\linewidth , height=5cm]{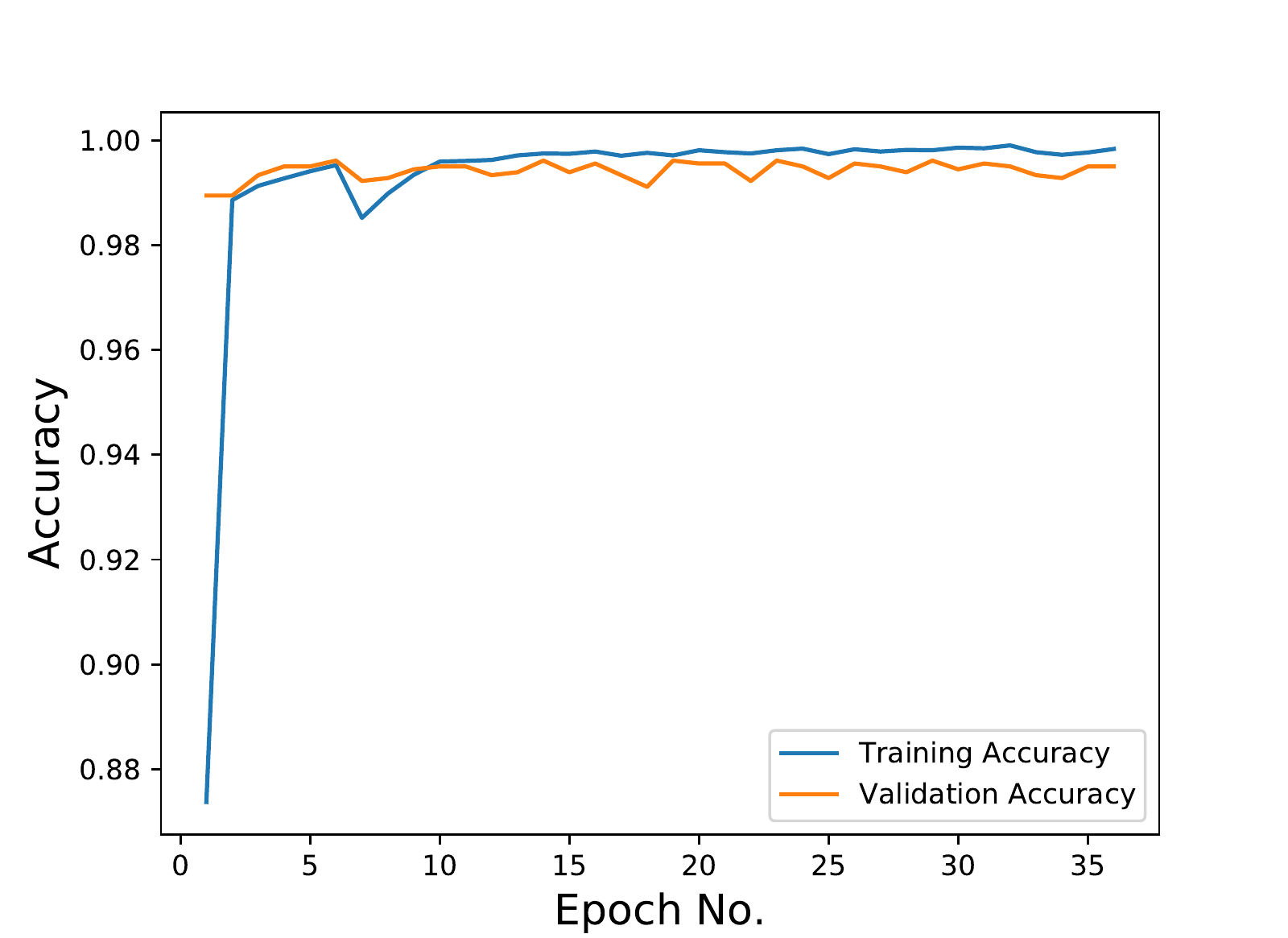}  \quad
  \caption{}
\end{subfigure}
\caption{Training and validation accuracies plotted as a function of the number of epochs for some typical cases---(a) proton-$\gamma$ and (b) helium-$\gamma$.}
\label{fig:fig_binary}
\end{figure}

To cross check the stability and robustness of our training procedure it is useful to examine how the accuracy score evolves during the run. A few typical plots displaying the evolution of the accuracy scores for both the training and validation sets as a function of the epoch number are shown in Fig.~\ref{fig:fig_binary} for proton-$\gamma$ and He-$\gamma$ classification. From the figure, we can clearly see that the accuracies saturate to an optimum value in an almost smooth fashion indicating good convergence of our machine parameters.

Finally, once we have trained the machine, we can now run it over the test data set and obtain the accuracy score as a quantification of the machine performance. The accuracy scores for the training, validation, and testing runs are tabulated in Table~\ref{Binary_table}, for different choices of the primary particle pairs, whose shower images we would like to distinguish. The training and validation accuracies listed in the table are for the optimized machine parameters selected after training.

\begin{table}
\centering
\begin{tabular}{|m{4cm}| m{2cm}| m{2cm}| m{2cm}| }
\hline
\multirow{2}{4em}{Classification}&\multicolumn{3}{c|}{Accuracy}\\
\cline{2-4}
 &  Training & Validation & Testing\\
\hline
$\gamma-$proton   &     0.997         &       0.996       &       0.991\\
$\gamma-$helium   &     0.995         &       0.996       &       0.997\\
$\gamma-$carbon   &     0.998         &       0.999       &       0.998\\
proton-helium          &     0.787        &        0.764       &       0.781          \\
proton-carbon            &     0.967        &        0.948       &       0.934          \\
helium-carbon                &     0.856        &        0.842       &       0.847       \\

\hline
\end{tabular}
\caption{Training, validation, and testing set accuracy scores for different pairs of SM primaries using our trained binary classifiers.}
\label{Binary_table}
\end{table}

From the accuracy scores in Table~\ref{Binary_table}, we see that our CNN binary classifier is able to discriminate between any two categories of CR shower images very competently. In particular, $\gamma$ initiated shower image patterns are very well discriminated from any hadron initiated shower by our CNN model, with accuracy scores greater than 99\% on training data. This number can be compared to other deep learning based discrimination methods that have been proposed in the literature. For example, ref.~\cite{2019APh...105...44S} found a $96$\% accuracy score for gamma-proton separation at H.E.S.S., which is an improvement over the standard H.E.S.S. BDT analysis based on Hillas parameters. Ref.~\cite{2019JPhCS1181a2048P} found a quality factor $Q =\epsilon_s/\sqrt{\epsilon_b}\simeq 2.99$, for gamma-proton separation at the TAIGA-IACT, where $\epsilon_s$ is the signal (gamma) acceptance, and $\epsilon_b$ is the background (proton) acceptance. We find a quality factor $Q\simeq 9.9$ for our binary classifier. Although on face value our results seem better than those presented in these previous works, we caution that a direct comparison between the results of these studies and our classifier would require a more detailed investigation since these studies use a broader energy range and broader incidence angles for their training and testing data, moreover in the case of the study in ref.~\cite{2019JPhCS1181a2048P}, the simulation is also for a different experiment. Additionally, we have restricted our analysis to the highest-quality 4-telescope data while other works like ref.~\cite{Shilon_2019} also consider events that triggered fewer telescopes, which are generally more difficult to classify.

Another interesting feature that we can see from Table~\ref{Binary_table} is that nuclei pairs with relatively close by atomic numbers --- such as proton-helium or helium-carbon --- have slightly lower accuracy scores (78\% and 85\% respectively). In contrast, CR showers initiated by nuclei that are further apart in atomic number (proton-carbon) yield much better accuracies for separation (93\%). This might be expected since at these energies, the primary interaction is between a nucleon in the charged cosmic ray primary with a nucleon in an atmospheric nucleus. The remaining nucleons in the the CCR are spectators to this interaction, although they contribute to the shower as a hadronic cascade. We might thus expect that the greater the number of spectator nucleons, the more distinct the shower pattern. However, it is difficult to separate shower images from different nuclei through a visual inspection, although the binary classifier seems to make this separation fairly well.

\subsubsection{Multi-category classification performance}
\label{sec:multi_result}

The goal for the multi-category classifier is to identify the categories of testing data which belong to one of multiple classes. As discussed earlier, it is similar to the binary classifier in terms of the machine architecture, except that it can work with more than two categories.

For the multi-category classifier, we first train our machine on labelled cosmic ray showers images initiated by all four different different types of SM primaries  belonging to our standard set, i.e. we select gamma and light nuclei ($\gamma$, $p$, He, C) primaries.

For our input data to the multi-category classifier, we use exactly the same split of the images in each class into training, validation, and test image sets, as in the case of the binary classifier. Thus, we take 10000 images of each category and split these into 8100 training images, 900 validation images, and 1000 testing images. The last 1000 images of each type are not seen by the classifier at any point during the training and so the performance metrics of the machine on these test images gives an accurate reflection of the discrimination capability among the different categories.

The training process is once again similar to that of the binary classifier. We use the mini-batch gradient descent method to optimize the machine parameters with a batch size of 100. The multi-category accuracy score is computed for the training and validation data sets after every training epoch. We use the early stop criterion as before to avoid over training.

Once the classifier is trained, we run on the test data set and compute the confusion matrix. We show the resulting confusion matrix in Table~\ref{Conf_mat}. As described in Sec.~\ref{sec:class_metrics}, this confusion matrix fully represents the performance of the machine. The diagonal entries of the matrix describe the ``true positives'', i.e. the instances that are correctly categorized (out of 1000 testing images of each type). The off-diagonal entries indicate the number of mis-classified images. For example, from the last row, labelled `carbon' in Table~\ref{conf_multi}, we conclude that out of 1000 testing carbon shower images, 871 are correctly tagged as carbon shower images whereas 125 and 4 of them are incorrectly labelled as helium and proton shower images, respectively.

We note two interesting features of the resulting confusion matrix. First, gammas are unlikely to be confused with anything other than $p$ images, and that too relatively rarely. Second, the classifier has significant difficulty separating proton and helium images, similar to what we have seen with the binary classifier.

\begin{table}
  \centering
  \begin{tabular}{|c|c|c|c|c|c|}
    \hline
    \multicolumn{2}{|c|}{} &\multicolumn{4}{|c|}{Predicted Labels} \\
     \cline{3-6}
     \multicolumn{2}{|c|}{}  & $\gamma$ & proton& helium & carbon  \\\hline
  \multirow{4}{*}{\rotatebox{90}{\makecell{Actual \\ Labels}}}  & $\gamma$ & \(985\) &  \(15\)&  \(0\) &  \(0\) \\
    & proton & \(4\) &  \(764\) &  \(208\) &  \(24\)  \\
   & helium & \(0\) &  \(231\) &  \(564\) &  \(205\)  \\
   & carbon & \(0\)&   \(4\)&   \(125\)    & \(871\)  \\
   \hline
  \end{tabular}
\caption{The confusion matrix for $\gamma$-proton-helium-carbon classification computed on the testing data set. The confusion matrix is defined in Sec.~\ref{sec:class_metrics}. Note, there are 1000 shower images for each category in the testing set.}
\label{conf_multi}
\end{table}

Based on the above confusion matrix we may compute the various simplified classification metrics for precision, recall, and $f1$ -score that we had defined in Eqs.~\ref{eq:prec},~\ref{eq:rec},~and~\ref{eq:f1}. These metrics for the multi-category classification are shown in Table~\ref{table_SM}.

As a reminder, for a given category: precision is a measure of how likely the classification reported for that category by the machine is likely to be correct; recall is a measure of how often images from a certain category are correctly classified into that category; and $f1$-score is the harmonic mean of the two.

\begin{table}
\centering
\begin{tabular}{|m{3cm} | m{2cm} | m{2cm} | m{2cm} |}
\hline
             & precision &   recall & f1-score \\ \hline
    $\gamma$  &  0.996 &  0.985  & 0.990   \\
    proton  & 0.753  & 0.764  & 0.759    \\
      helium  &  0.629 &  0.564 &  0.595   \\
     carbon  &  0.792 &  0.871 &  0.830      \\
     \hline
     \hline
   weighted avg   & 0.792 &  0.796 &  0.793    \\
\hline
\end{tabular}
\caption{Performance metrics for $\gamma$-proton-helium-carbon classification. The performance metrics are defined in Sec.~\ref{sec:class_metrics}.
}
\label{table_SM}
\end{table}



From our table, we can see that precision and recall are highest for $\gamma$ $\sim$ 99\%, and worse for charged CRs. For C~nuclei we find a relatively high recall score of 87\%, since these nuclei are unlikely to be mistaken for other nuclei that we have considered. However, the precision for C~nuclei is much poorer at 79\%, and this is because He nuclei can often be mistaken for C~nuclei by our classifier. As mentioned earlier protons are often mislabelled as He and vice-versa, leading to lower precision and recall scores for these nuclei. We have also reported the weighted-average of each of these scores (averaged over all categories) in our table.

As one would expect, the multi-category classification metrics are more modest than that of binary classification, since there is more potential for mislabelling of particular images. Nevertheless, in absolute terms the performance is good, especially for $\gamma$-nuclei separation and $p$-C or He-C separation. These results for multi-category classification also align well with our expectation based on binary classification of the shower images, for e.g. in terms of $p$-He being harder to separate.

\subsubsection{Classification with other energies and angles for the primary}
\label{otherenergies}

In order to check the robustness of the CNN classifier methodology, we have also performed binary and multi-category classification for other combinations of energy bins (100~TeV and 60~TeV) and zenith angles ($0^\circ$ and $45^\circ$). The result of this classification is shown in appendix \ref{appendix_classification}. The results for the other energy bins and zenith angles are almost similar to the result that we have discussed in this section for the $E=100$~TeV, $\theta_0= 0^\circ$ case. This enhances confidence in the power of the CNN strategy that we have adopted.

\subsubsection{Can the classification result be explained by the differences in event size?}
\label{ref_response}

It is well known that for a given primary energy, gamma-ray-initiated
showers produce \mbox{$\sim$~2-3} times the light output of proton-initiated showers. This would lead to larger event sizes (where we define `size' as the total photo-electron counts summed over all pixels in all the four detectors) for gamma ray showers as compared to hadronic showers of the same energy.

Since the energy range that we have allowed for the primaries is narrow, between $99.5-100.5$~TeV, one obvious concern might be that the binary and multi-category classifiers that we have constructed may have mainly learnt about the event size and used it as a discriminatory variable. Such a discriminant would not be as effective at separating gamma-hadron showers in realistic experimental data where the separation must be achieved for primaries which span a wider range of energies. In the realistic scenario, shower shape variables, rather than size must be used as the primary discriminant between the different types of showers, although size maybe an important secondary discriminant.

\begin{figure}
  \centering
  \includegraphics[scale=0.6]{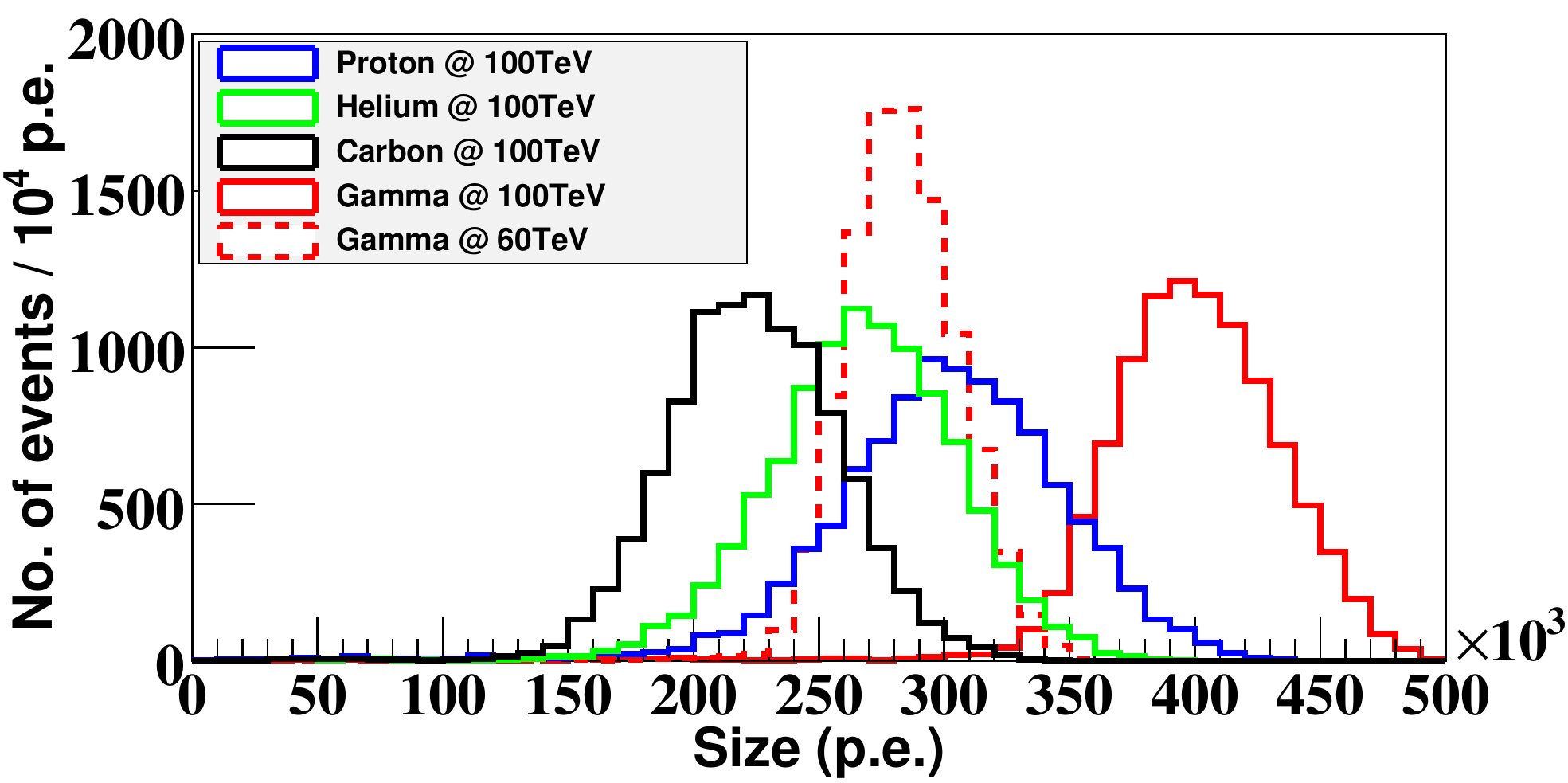}

\caption{The event size distribution of 100~TeV $\gamma$ shower images and 100 TeV hadronic shower images when zenith angle, $\theta_0$ is $0^\circ$ are significantly different. However, the size distribution of 60~TeV $\gamma$ shower images and 100 TeV hadronic shower images is not so different and thus event size is not a good enough discriminatory variable.}
  \label{fig:sizedist}
\end{figure}

The qualitative difference in event size can be seen by looking at the relative brightness of detector images of proton and gamma showers in Fig.~\ref{TelImage}. More quantitatively, we can plot the distributions of event size for all the images in our sample at 100 TeV, $0^\circ$, for different primary species. These distributions are shown in Fig.~\ref{fig:sizedist}. We can see quite clearly that proton showers, as well as showers initiated by He and C, have a smaller event size than gammas of the same energy. Thus, even without the use of ML techniques, one could place a cut on the total event size and achieve very good discrimination between hadron and gamma initiated showers at 100~TeV.

Now, we would like to show that the discrimination ability of our binary and multi-category classifiers can not simply be attributed to a difference in event sizes. To see this, consider the problem of separating 60~TeV gamma ray showers from 100 TeV hadronic showers. In Fig.~\ref{fig:sizedist}, we have also plotted the size distribution for 60 TeV gamma ray showers. As can be seen from the figure, the size distribution for such showers is similar to that of 100 TeV hadronic showers. Thus, if we can achieve a similar discrimination ability between gammas with this lower energy and our 100 TeV hadronic sample, it would demonstrate that our classifiers can learn some other discriminatory variables which characterize the shower, other than just the event size.

We have repeated our analysis of Sec.~\ref{sec:binary_result} and Sec.~\ref{sec:multi_result}, for the binary and multi-category classifiers, with the use of a set of 10000 60 TeV gamma ray shower images instead of the 100 TeV gamma ray showers that we had previously considered, while keeping the hadronic 100~TeV image set the same. The results that we obtain for this analysis for the binary classifier are shown in Table~\ref{bin_0Deg_G60H100} and for the multi-category classifier in Table~\ref{ConfMat:0Deg}.

\begin{table}
\centering
\begin{tabular}{|m{4cm}| m{2cm}| m{2cm}| m{2cm}| }
\hline
\multirow{2}{4em}{Classification}&\multicolumn{3}{c|}{Accuracy}\\
\cline{2-4}
 &  Training & Validation & Testing\\
\hline
$\gamma-$proton   &     0.999         &       0.998      &       0.994\\
$\gamma-$helium   &     0.999         &       0.998       &       0.997\\
$\gamma-$carbon   &     0.999         &       1.000       &       0.998\\

\hline
\end{tabular}
\caption{Training, validation, and testing set accuracy scores for different pairs of SM primaries by training and testing our classifier on a mixed set of 100 TeV hadronic shower images and 60 TeV gamma shower images (zenith angle of shower, $\theta_0$ is $0^\circ$).}
\label{bin_0Deg_G60H100}
\end{table}

\begin{table}
\centering
\begin{tabular}{|m{3cm} | m{2cm} | m{2cm} | m{2cm} |}
\hline
             & precision &   recall & f1-score \\ \hline
    $\gamma$  &  0.993 &  0.995  & 0.994   \\
    proton  & 0.771  & 0.746  & 0.758    \\
      helium  &  0.644 &  0.595 &  0.618   \\
     carbon  &  0.799 &  0.885 &  0.840      \\
     \hline
     \hline
   weighted avg   & 0.802 &  0.805 &  0.803    \\
\hline
\end{tabular}
\caption{Performance metrics for multi-category classification computed after training and testing data on a mixed set containing 100 TeV hadronic shower images and 60 TeV $\gamma$ shower images (zenith angle of shower, $\theta_0$ is $0^\circ$). }
\label{ConfMat:0Deg}
\end{table}

The results in these tables are similar to those that we obtained in the case where we used samples of 100~TeV hadron and 100~TeV gamma shower images (compare Table~\ref{bin_0Deg_G60H100} with Table~\ref{Binary_table} and Table~\ref{ConfMat:0Deg} with Table~\ref{table_SM}, respectively). Thus, even when event size can not be used as a good discriminator of gammas and hadrons, our binary and multi-category classifiers still show excellent gamma-hadron separation ability indicating that the machines are learning more subtle features of the data such as the shape of the shower.

\subsection{Unsupervised learning and anomaly detection}
\label{sec:result_anomaly}

We now come to the anomaly finder part of our study. The basic question we are trying to address is this: Given some typical images corresponding to showers initiated by SM primaries, can we train a machine to learn features of these images in such a way that it is able to flag anomalous events that it has never previously encountered, and which have features which are different from the training data set?

The advantage of such a machine compared to a binary or multi-category classifier is that it would be model agnostic as to the features of new BSM physics that might be seen at a cosmic ray experiment. The ability to flag anomalies does not have to do with specific  features of the anomaly, but rather the inability of the anomalous events to conform to expectations of the SM shower images.

We use an autoencoder which attempts to learn features of training images. The architecture of the autoencoder has already been discussed in Sec.~\ref{sec:ae_arch}. The autoencoder is trained on our standard image set of showers initiated by $\gamma$, $p$, He, and C nuclei. For the input to the autoencoder, we use the remapped images with a $\sqrt{x}$ remapping, as described in Sec.~\ref{sec:remap}.

For each type of SM primary we take 10000 remapped images that we have generated, where the primary has an energy centered around 100~TeV and a zenith and azimuthal angle both of $0^\circ$. For each SM primary, the data is split in to 8100 training images, 900 validation images, and 1000 testing images. We refer to the collected images for all primaries as the ``training set'', ``validation set'', and ``standard test set'', respectively.

For testing our anomaly finder, we also construct 4000 remapped images of a prototypical BSM shower initiated by a $Z^\prime\rightarrow e^+e^-$ with the same energy, zenith and azimuthal angle as the SM primaries. We refer to these images as the ``anomalous test set''. All the $Z^\prime$ images, as well as the SM images that are reserved for testing, are only used at the testing stage and are not seen by the machine during the training phase.

We also present our results for the anomaly finder for other choices of energies and angles in appendix~\ref{appendix_anomaly}.


\begin{figure}
\begin{subfigure}{0.45\textwidth}
\hspace{-1.2 cm}
\includegraphics[height=6cm, width=9cm]{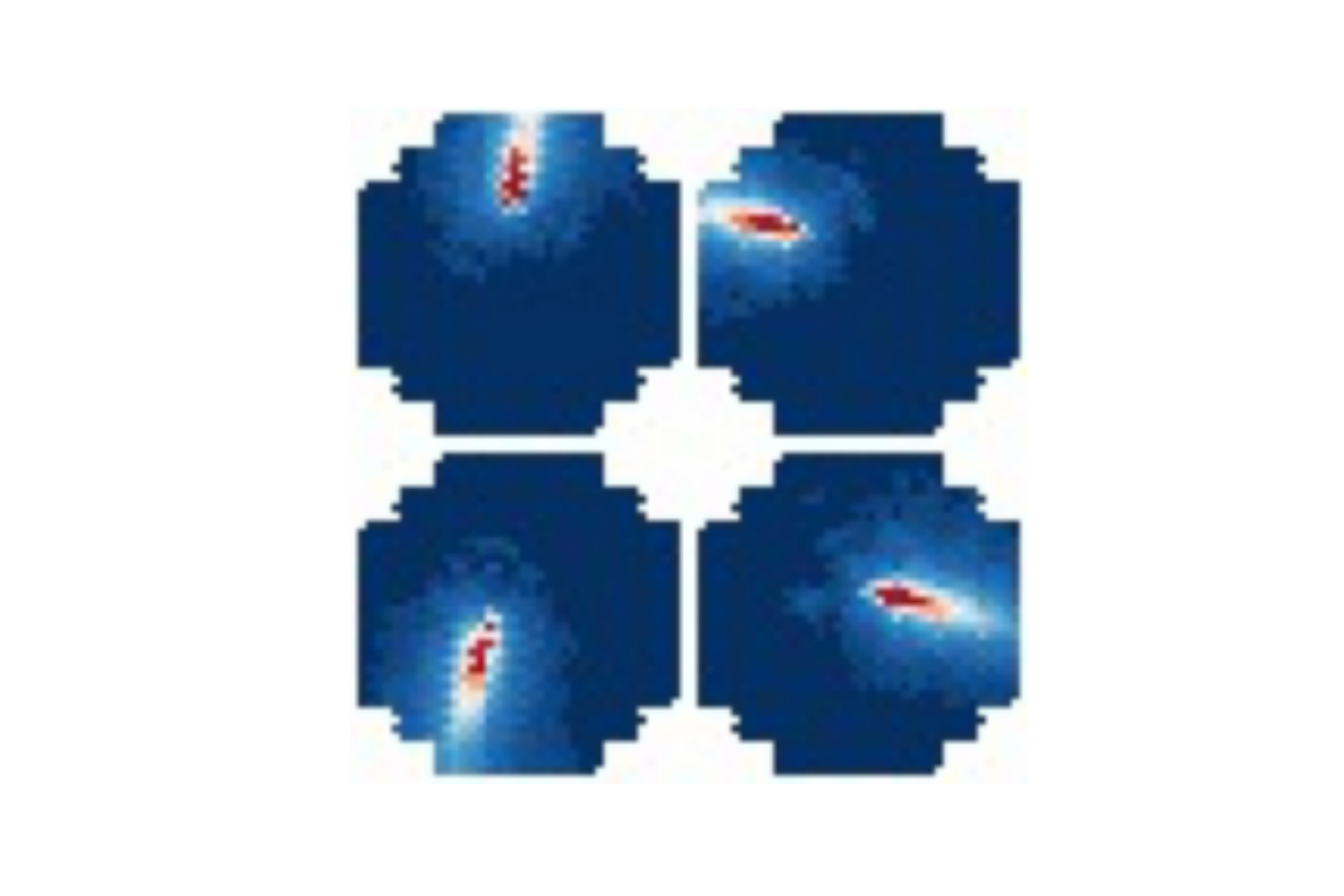}
\subcaption{}
\end{subfigure}
\begin{subfigure}{0.45\textwidth}
\hspace{-1.2cm}
\includegraphics[height=6cm, width=9cm]{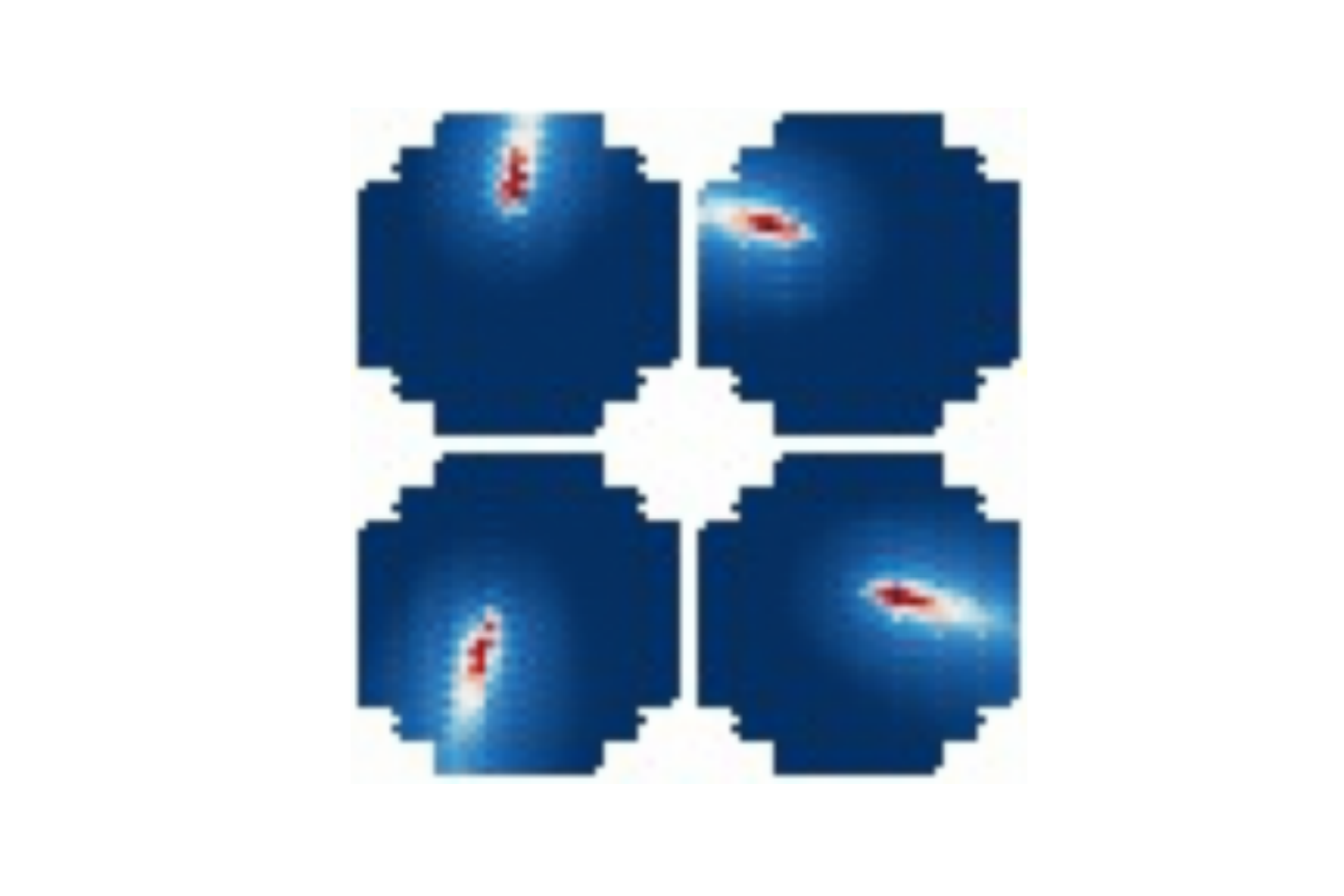}
\subcaption{}
\end{subfigure}
\caption{(a) Original (remapped) proton image passed as input to the trained autoencoder, (b) reconstructed proton image obtained as output from the autoencoder. The reconstructed image seems to have captured all the discernible features of the input image. Because of only slight differences between the input and output images, we will obtain a small mean-squared error between the two. Note that we use remapped images with a $\sqrt{x}$ rescaling (see Sec.~\ref{sec:remap}) as inputs to the autoencoder for both training and testing purposes. }
\label{proton_reco}
\end{figure}
\begin{figure}[H]
\begin{subfigure}{0.45\textwidth}
\hspace{-1.2cm}
\includegraphics[height=6cm, width=9cm]{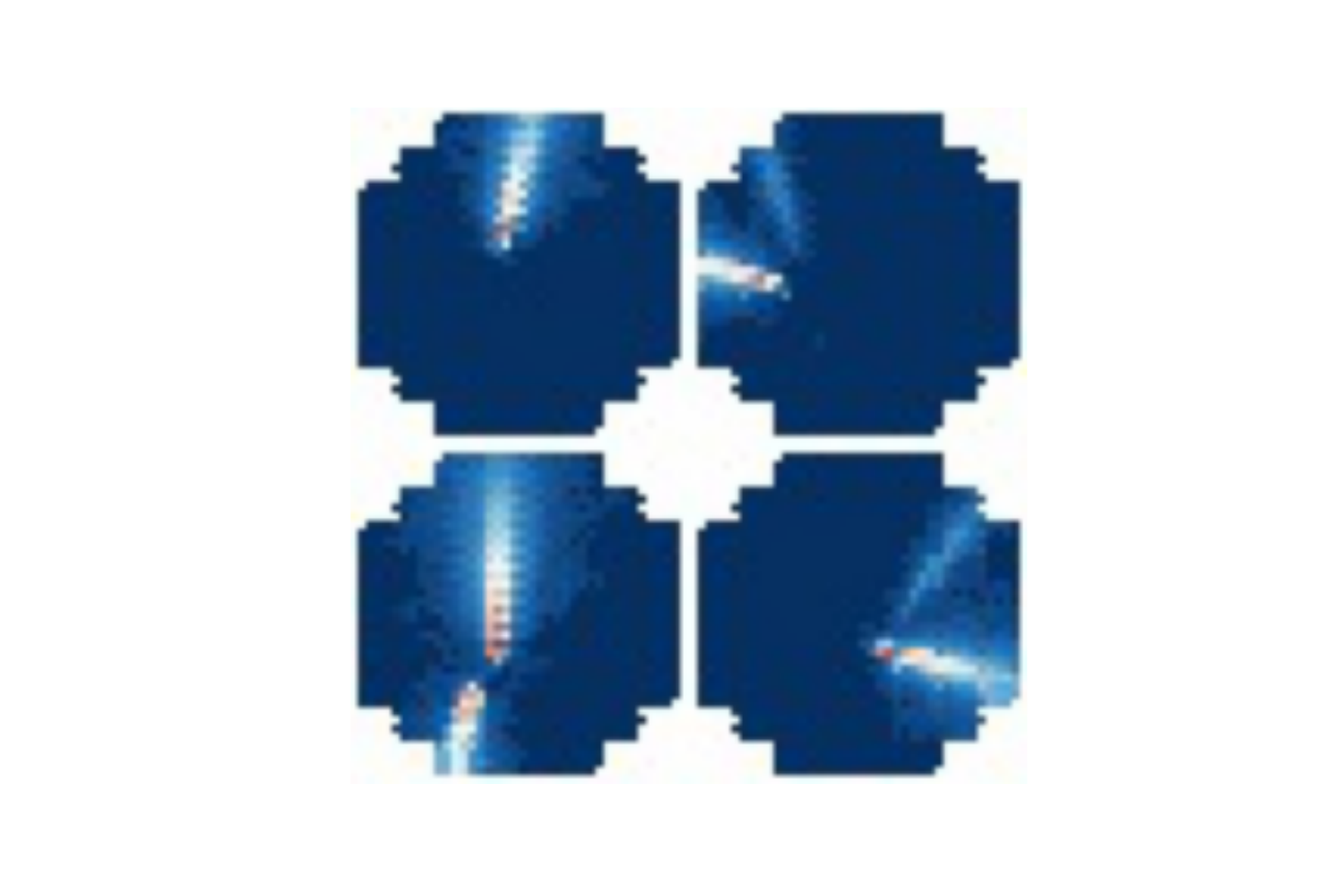}
\subcaption{}
\end{subfigure}
\begin{subfigure}{0.45\textwidth}
\hspace{-1.2cm}
\includegraphics[height=6cm, width=9cm]{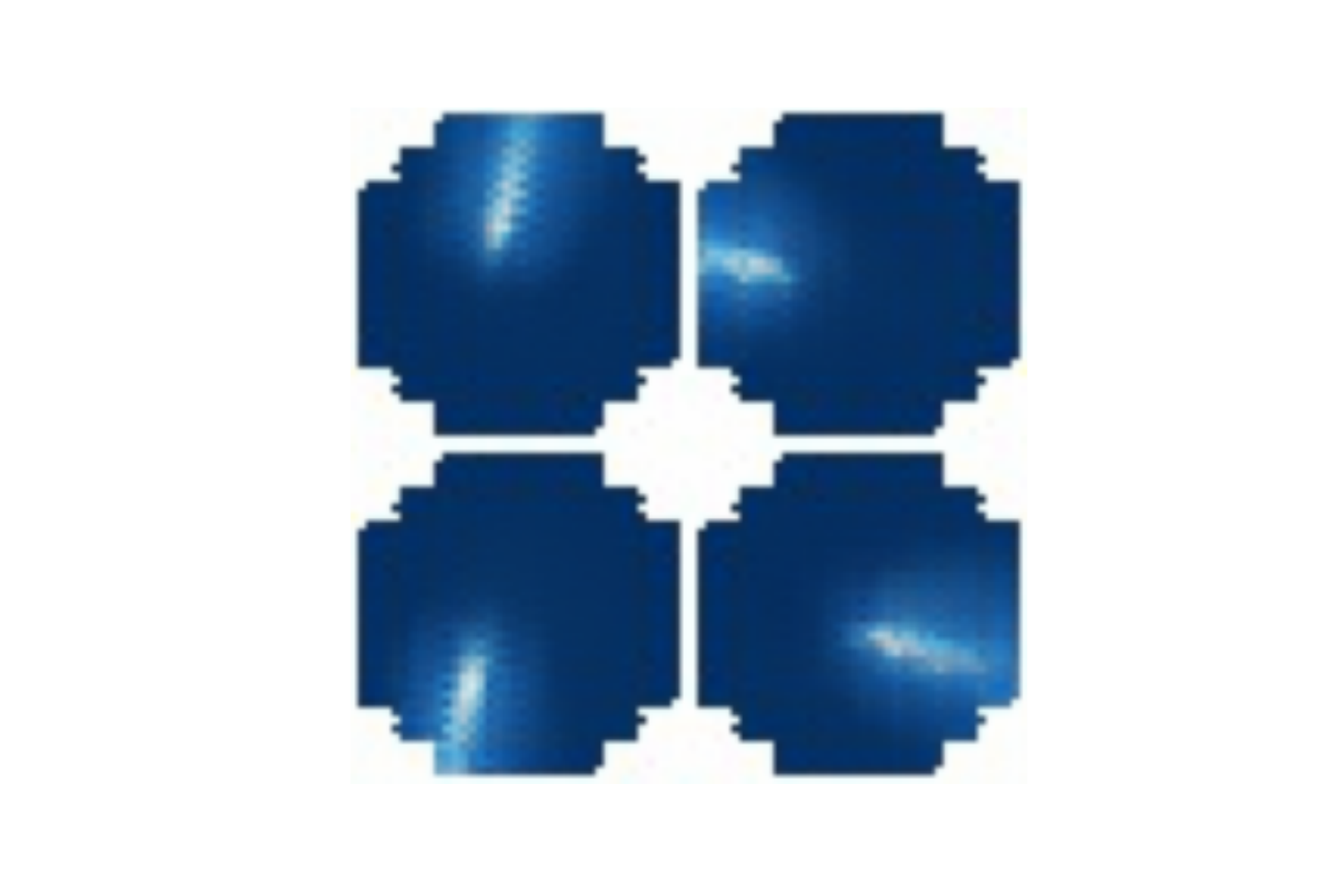}
\subcaption{}
\end{subfigure}
\caption{(a) Original (remapped) $Z^\prime$ image passed as input to the trained autoencoder, (b) reconstructed Z$^\prime$ image obtained as the output of the autoencoder. The $Z^\prime$ is  anomalous because such events have not been seen by the autoencoder during the training phase. The reconstructed image does not capture the fainter second prong of the input Z$^\prime$ image. The difference between the input and output images will lead to a large mean-squared error. Note that we use remapped images with a $\sqrt{x}$ rescaling (see Sec.~\ref{sec:remap}) as inputs to the autoencoder. }
\label{Zprime_reco}
\end{figure}

The machine learning is unsupervised, because we do not label the input training data to the machine, it simply learns features of all the inputs and attempts to reconstruct the images as accurately as possible from a compressed representation of the original images.

Similar to our classifiers, we once again use the mini-batch gradient descent method to train the autoencoder, with batches of 100 randomly selected events from the training set. For each batch, the machine calculates a total loss function (which is the mean-squared error or MSE, see Eq.~\ref{mse}) and then updates the learnable parameters in at attempt to minimize the loss. This process is repeated until all the events in the training data are processed. This entire set of steps constitutes one epoch. We then compute the MSE for the validation set at the end of the epoch.

This process is continued until the validation set MSE does not decrease below the MSE of an epoch number $i_\textrm{crit}$ for 30 more consecutive epochs (`early stop' criterion). We then take the final machine parameters to be those of the epoch $i_\textrm{crit}$ which has the smallest validation set MSE.  After this the machine is trained and we no longer change the learnable parameters. The machine is now ready for testing to evaluate its performance as an anomaly finder.

We now run the machine over the combined test set comprising of the standard test set and the anomalous test set.

\begin{figure}[t]
\centering
\includegraphics[scale=0.7]{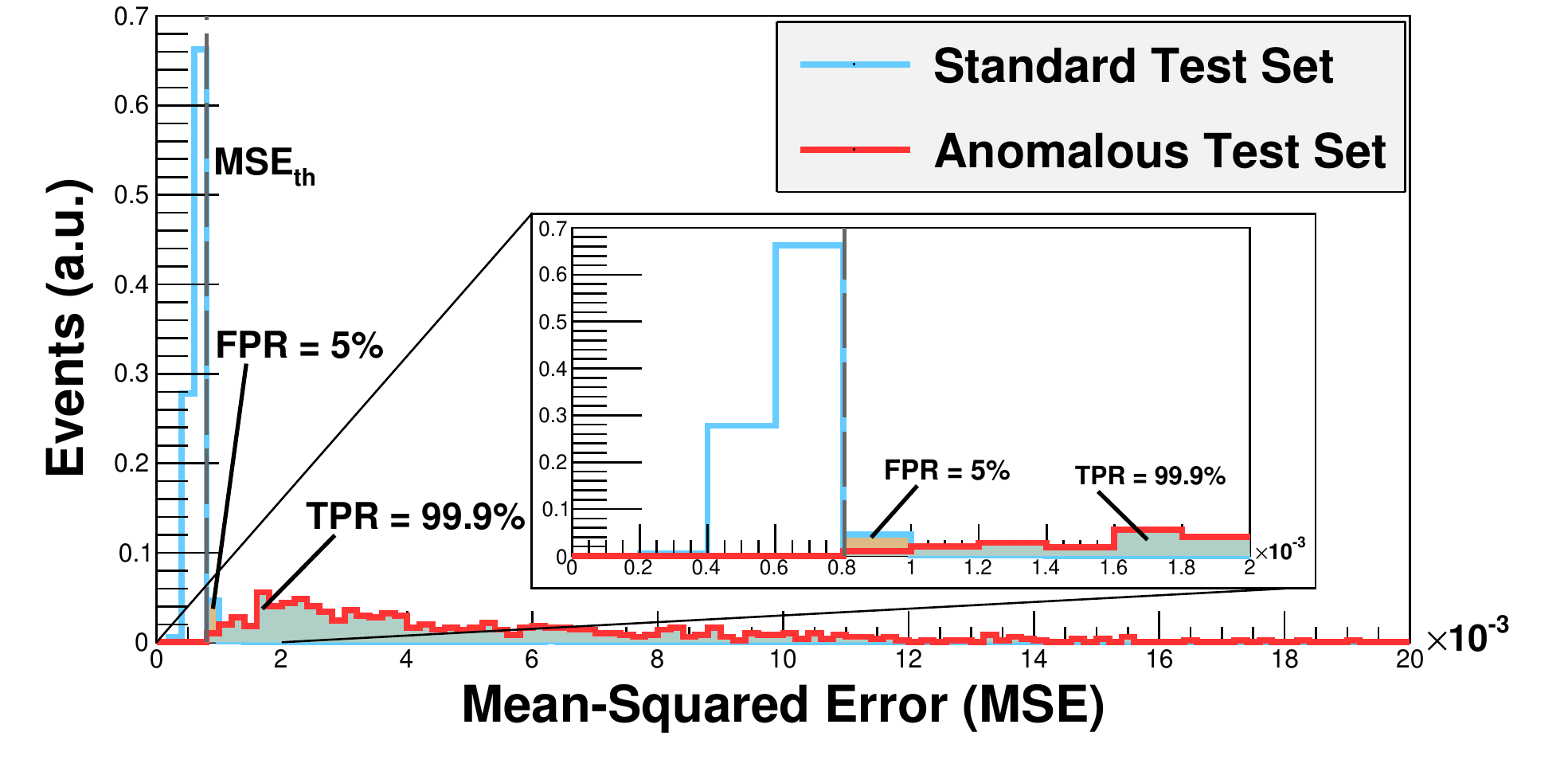}
\caption{Mean-squared error distribution for the standard (blue) and anomalous (red) test sets comprising of shower images initiated by SM and $Z^\prime$, respectively. Since the autoencoder has been trained on SM type images, the standard test set has relatively low reconstruction errors compared to the $Z^\prime$ images, which have much larger reconstruction errors on average. Also shown is a threshold MSE cut (vertical grey-dashed). Images with MSE larger than this threshold are classified as anomalous. The fractional area under the red curve to the right of the threshold gives the true positive rate (TPR) for anomalies, whereas the fractional area under the blue curve to the right of the threshold gives the false positive rate (FPR). Here, we have chosen MSE$_\textrm{th}$ so that the FPR = 5\%.}
\label{mse_Z}
\end{figure}

Before looking at aggregate data for the entire test data set, it is useful to get an intuitive feel for the performance of the anomaly finder on individual images in the test set. The autoencoder should have learnt essential features of the training data which was composed of images initiated by SM primaries. Therefore, when fed a standard test image as input, the trained autoencoder should reconstruct the original image nearly faithfully. However, if the input to the machine is a $Z^\prime$ initiated anomalous shower image, the reconstruction of the image should go awry since the autoencoder will not be able to capture all the features of the $Z^\prime$ in the compressed bottleneck layer.

In Figs.~\ref{proton_reco} and ~\ref{Zprime_reco}, we show two representative examples of the image reconstruction from the autoencoder based on a proton images from the standard test set, and a $Z^\prime$ image from the anomalous test set. Qualitatively, already a crude comparison-by-eye tells us that the reconstruction of the image from the standard test set is better than that of the shower image from the anomalous test set. The proton image has only a single prong feature which is well reconstructed. However, for the $Z^\prime$ image we see that the fainter second prong is missed in the reconstructed image. We observe a similar trend when looking at other reconstructed shower images from the test sets as well. The qualitative comparison is already encouraging and suggests that the trained machine seems to be good at learning features of the SM data, but the $Z^\prime$ images are sufficiently distinct, so that their reconstruction is poor.

\begin{figure}[t]
\begin{center}
\includegraphics[height=11cm,width=13cm]{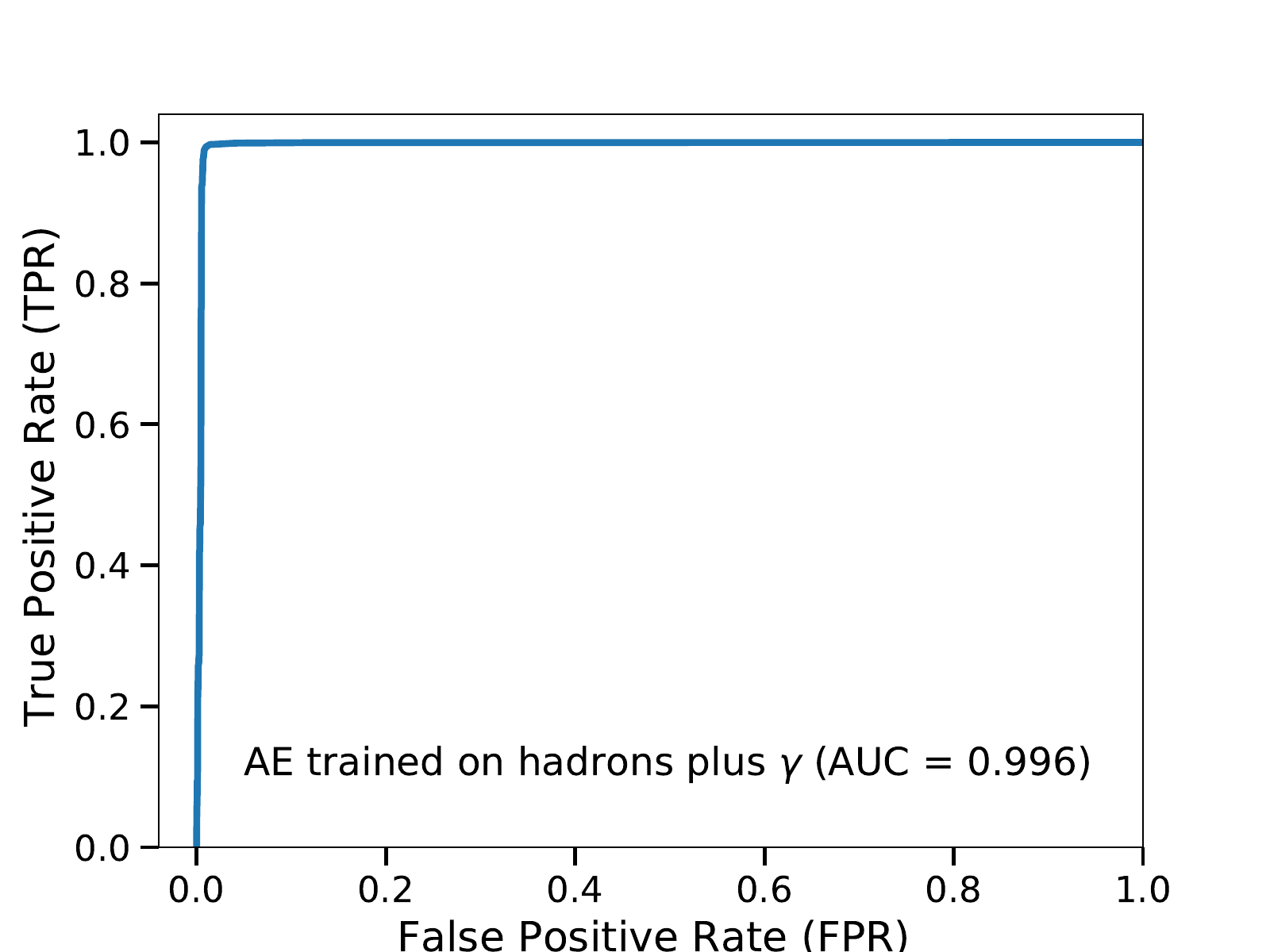}
\end{center}
\caption{ROC curve showing a comparison between the TPR and FPR as the MSE$_\textrm{th}$ is varied. This ROC curve is constructed using the MSEs computed on the standard and anomalous test sets, whose distributions are shown in Fig.~\ref{mse_Z}. The area-under-the ROC curve (AUC) is 0.996. The shower images used to construct this ROC curve are initiated by particles with an energy of 100 TeV and zenith and azimuthal angles selected in a cone of 1.5$^\circ$ around $\theta_0 = 0^\circ$ and $\phi_0 = 0^\circ$.
}
\label{ROC}
\end{figure}

To sharpen the above qualitative observations, we may get a quantitative estimate of how our autoencoder performs as an anomaly detector by looking at the distribution of MSE values for the entire standard test set and the anomalous test set. These normalized distributions are shown in Fig.~\ref{mse_Z}. We see from the figure that the MSE values for the standard test set are on average lower than those of the anomalous test set, indicating that SM initiated shower images have low reconstruction errors. Moreover, the MSE distribution of the standard and anomalous test sets are fairly well separated. We can therefore select a threshold MSE value, MSE$_\textrm{th}$ such that images which have a MSE greater than MSE$_\textrm{th}$ are classified as anomalous, and images with MSE below this threshold are classified as standard. The choice of MSE$_\textrm{th}$ is arbitrary, but for different choices of this threshold, we would obtain different selection efficiencies for tagging events as standard or anomalous.

We can now plot an ROC curve (described in Sec.~\ref{mse_ROC_AUC}) for the efficiency of tagging anomalous events as anomalous (True Positive Rate) versus the efficiency of tagging standard events as anomalous (False Positive Rate), as we change MSE$_\textrm{th}$. The ROC curve corresponding to the MSE distributions in Fig.~\ref{mse_Z} is displayed in Fig.~\ref{ROC}. The AUC corresponding to this ROC curve is 0.996. At the benchmark FPRs of 10\%, 5\%, and 1\%, we find TPRs of 99.95\%, 99.9\%, and 99.4\%, respectively. This indicates that the anomaly finder can flag the $\zp$ type anomalous events with very high confidence while maintaining a low false alarm rate. By choosing a stringent value of MSE$_\textrm{th}$, such that the FPR is 1\%,  we thus expect to be able to enhance the signal-to-background ratio for anomalous events by a factor of nearly 100:1. This could potentially be increased further, but the statistics of our simulated events are insufficient to reliably understand the ROC curve at lower FPRs such as at 0.1\% or lower.

The ROC curves for other energy and zenith angle combinations are similar. These ROC curves are shown in appendix~\ref{appendix_anomaly}. Hadronic shower images, compared to gamma shower images, can contain much
greater variation including having multiple clusters, so one might expect greater confusion
with the $\zp$ showers for the hadronic images as opposed to the gamma images. To check this, we have also repeated the tests for our autoencoder by training on shower images of only hadrons, and presented the resulting ROC curve in appendix~\ref{appendix_anomaly}. In all cases we find that the ROC curves are similar to that of Fig.~\ref{ROC}, which increases our confidence in the robustness of our anomaly finder.

\section{Conclusion}
\label{sec:Conclusions}

The study of very high energy gamma rays and charged cosmic rays gives us a window into physics at energy scales beyond the reach of present day collider experiments. In this work, our focus was on Imaging Air Cherenkov Telescopes which have been employed with great effect to search for VHE gamma rays. We started by posing three problems relevant to IACTs: 1) the separation of gamma and hadron showers, 2) the identification of various hadronic primaries in cosmic ray showers, and 3) the identification of anomalous events at IACTs that do not conform to known shower patterns of either hadronic or gamma ray primaries. The latter two problems have been relatively less explored in the literature.

In our work, we addressed these problems using the approach of Deep Learning. The first two problems of gamma-hadron separation and classification of hadronic primaries are well suited to a supervised learning approach. We built a binary and a multi-category classifier using a convolutional neural network. We found that our classifiers can separate gamma rays from protons with $>99\%$ accuracy, which on face value is better than results found elsewhere in the literature. However, as we have cautioned, a detailed comparison study would be needed to establish a definitive claim about the relative efficacies of the ML approaches adopted here and in other works. We are also able to achieve good but relatively modest performance for identification of nuclear species, with best identification of carbon nuclei initiated showers. We found that proton and helium nuclei initiated showers are relatively harder to differentiate.

In order to identify anomalous events at IACTs, we presented a design of an autoencoder architecture which is similar to those suggested for use at collider physics experiments for a similar purpose. The machine was trained on shower images of purely SM gamma/hadron initiated cosmic ray shower images. When testing our machine, we focused on the prototypical case of a BSM $\zp$ decaying to $e^+e^-$, while remaining agnostic as to the source of such a $\zp$. We found that our autoencoder could increase the signal-to-background ratio by a factor of $\sim$~100 with 99\% background rejection, however more expensive simulations with greater statistics are needed to identify the potential for separation at even higher background rejection rates. Although a dedicated BSM search would undoubtedly perform better, such strategies are model dependent, whereas the power of the anomaly finder lies in its model independent discrimination ability. This tool thus allows us to utilize hitherto untapped potential of IACTs.

Our study made use of full cosmic ray shower images at IACTs and thus used the full detector information. This is unlike studies which work with reduced information such as Hillas parameters. In addition, simulating BSM events at cosmic ray experiments is complicated given the publicly available tools, and we have highlighted some of these difficulties when discussing our $\zp$ simulations.

We hope that our study has demonstrated the power of ML techniques for experimentalists working on the analysis of current and future IACT data, and in particular for the upcoming CTA. Given the extraordinary energy reach of these experiments, it would be prudent to employ ML tools like our autoencoder to search for BSM physics in a model independent way. For model dependent studies, further developments in simulation tools for BSM physics at cosmic ray experiments are needed in order to exploit the full power of the data that is expected from these experiments in the future.

Although our studies were based on IACTs which have traditionally been used for gamma ray searches, we expect that since our techniques our based on image pattern recognition, that they can easily be ported to other cosmic ray experiments employing different detection techniques.

\acknowledgments
We acknowledge useful discussions related to \texttt{CORSIKA} with Dieter Heck, Ralf Ulrich and Tanguy Pierog over emails. We are thankful to Konrad Bernl\"ohr for helping us with the functioning of \texttt{sim\_telarray}. We owe thanks to Stefan Ohm and Maximilian N\"othe for helping us with useful information regarding the H.E.S.S. experiment and \texttt{ctapipe} respectively. We also thank Pratik Majumdar for informative discussions. We thank Sayan Saha for initial collaboration during the early stages of this work. AT acknowledges support from an Early Career Research award, from the Department of Science and Technology, Government of India.


\appendix

\section{Classification result for different energy bins and zenith angles}
\label{appendix_classification}
We have repeated the simulations for our binary and multi-category classifiers with Standard Model initiated cosmic rays showers with other choices of energies and zenith angles for the primary. We have taken the following combinations of energy $E$ and zenith angle $\theta_0$ - (100~TeV, 0$^\circ$), (100~TeV, 45$^\circ$), (60~TeV, 0$^\circ$), and (60~TeV, 45$^\circ$). We continue to fix the azimuthal angle $\phi_0 = 0^\circ$ in all cases. The results for (100~TeV, 0$^\circ$) were presented in the main text. In this appendix we show the results for all the other combinations.

We note here a couple of comments about these results. First, the energies are chosen using a power law distribution described in Sec.~\ref{sec:EAS_sim} in a 1 TeV region around $E$. Second, the zenith and azimuthal angles for the cosmic ray shower axes, are chosen in a cone of semi-vertical angle $1.5^\circ$ around $\theta_0$ and $\phi_0$. Third, the telescope angles $\theta_\textrm{tel}$, $\phi_\textrm{tel}$ are always set such that they point in the $\theta_0$ and $\phi_0$ direction. Fourth, for each energy/angle combination we generate 10000 events for each SM primary, which are split into 8100 training events, 900 validation events, and 1000 testing events. Importantly, we do not mix energy and angle combinations when training or testing. Thus, all SM primaries with a fixed energy and angle combination are used for training, validation, and testing. The procedure for training is as described in the main text.

\begin{table}[H]
\centering
\begin{tabular}{|m{4cm}| m{2cm}| m{2cm}| m{2cm}| }
\hline
\multirow{2}{4em}{Classification}&\multicolumn{3}{c|}{Accuracy}\\
\cline{2-4}
 &  Training & Validation & Testing\\
\hline
$\gamma-$proton   &     0.993         &       0.993       &       0.988\\
$\gamma-$helium   &     0.998         &       0.998       &       0.996\\
$\gamma-$carbon   &     0.999         &       0.999       &       0.999\\
proton-helium          &     0.761        &        0.762       &       0.754         \\
proton-carbon            &     0.959        &        0.938       &       0.945          \\
helium-carbon                &     0.867        &        0.824      &       0.821       \\
\hline
\end{tabular}
\caption{Training, validation, and testing set accuracies for different binary classifications among standard model initiated shower images resulting from showers with energies $E=100$~TeV and with zenith angle $\theta_0 = 45^\circ$.}
\label{Binary_table_100_45}
\end{table}

\begin{table}[H]
\centering
\begin{tabular}{|m{4cm}| m{2cm}| m{2cm}| m{2cm}| }
\hline
\multirow{2}{4em}{Classification}&\multicolumn{3}{c|}{Accuracy}\\
\cline{2-4}
 &  Training & Validation & Testing\\
\hline
$\gamma-$proton   &     0.997         &       0.994       &       0.994\\
$\gamma-$helium   &     0.999         &       0.998       &       0.999\\
$\gamma-$carbon   &     0.998         &       0.999       &       0.999\\
proton-helium          &     0.788        &        0.759       &       0.756          \\
proton-carbon            &     0.956        &        0.952       &       0.937          \\
helium-carbon               &     0.886        &        0.842       &       0.810       \\
\hline
\end{tabular}
\caption{Training, validation, and testing set accuracies for different binary classifications among standard model initiated shower images resulting from showers with energies $E=60$~TeV and with zenith angle $\theta_0 = 0^\circ$.}
\label{Binary_table_3}
\end{table}

\begin{table}[H]
\centering
\begin{tabular}{|m{4cm}| m{2cm}| m{2cm}| m{2cm}| }
\hline
\multirow{2}{4em}{Classification}&\multicolumn{3}{c|}{Accuracy}\\
\cline{2-4}
 &  Training & Validation & Testing\\
\hline
$\gamma-$proton   &     0.996         &       0.996       &       0.993\\
$\gamma-$helium   &     0.993         &       0.998       &       0.997\\
$\gamma-$carbon   &     0.935         &       1.000       &       0.999\\
proton-helium          &     0.773        &        0.762       &       0.763          \\
proton-carbon           &     0.948        &        0.938       &       0.941          \\
helium-carbon               &     0.827        &        0.816       &       0.820       \\
\hline
\end{tabular}
\caption{Training, validation, and testing set accuracies for different binary classifications among standard model initiated shower images resulting from showers with energies $E=60$~TeV and with zenith angle $\theta_0 = 45^\circ$.}
\label{Binary_table_4}
\end{table}

The results for the binary  classifier for the other energy and angle combinations are shown in tables~\ref{Binary_table_100_45}, \ref{Binary_table_3}, \ref{Binary_table_4}, and those for the multi-category classifier are shown in tables~\ref{table_SM_2}, \ref{table_SM_3}, \ref{table_SM_4}. The results and trends in these tables are similar to those for the (100~TeV, 0$^\circ$) combination discussed in the main text.

\begin{table}[H]
\centering
\begin{tabular}{|m{3cm} | m{2cm} | m{2cm} | m{2cm} |}
\hline
             & precision &   recall & f1-score \\ \hline

       $\gamma$  &  0.993 &  0.988  & 0.990   \\

      proton  & 0.781  & 0.728  & 0.754    \\
      helium  &  0.627 &  0.625 &  0.626   \\
      carbon  &  0.807 &  0.868 &  0.836   \\
    \hline
    \hline
   weighted avg   & 0.802 &  0.802 &  0.802\\
\hline
\end{tabular}
\caption{Precision, recall, and f1 scores for $\gamma$-proton-helium-carbon shower images with energies $E=100$~TeV and with zenith angle $\theta_0 = 45^\circ$.}
\label{table_SM_2}
\end{table}

\begin{table}[H]
\centering
\begin{tabular}{|m{3cm} | m{2cm} | m{2cm} | m{2cm} |}
\hline
             & precision &   recall & f1-score \\ \hline

       $\gamma$  &  0.989 &  0.999  & 0.994   \\
       proton  & 0.804  & 0.705  & 0.751    \\
      helium  &  0.592 &  0.614 &  0.603   \\
       carbon  &  0.773 &  0.832 &  0.802   \\

   \hline
   \hline
   weighted avg   & 0.790 &  0.788 &  0.787   \\
\hline
\end{tabular}
\caption{Precision, recall, and f1 scores for $\gamma$-proton-helium-carbon shower images with energies $E=60$~TeV and with zenith angle $\theta_0 = 0^\circ$.}
\label{table_SM_3}
\end{table}

\begin{table}[H]
\centering
\begin{tabular}{|m{3cm} | m{2cm} | m{2cm} | m{2cm} |}
\hline
             & precision &   recall & f1-score\\ \hline

       $\gamma$  &  0.994 &  0.997  & 0.996   \\
       proton  & 0.823  & 0.687  & 0.749   \\
      helium  &  0.625 &  0.674 &  0.648   \\
      carbon  &  0.801 &  0.868 &  0.833   \\
      \hline
\hline

   weighted avg   & 0.811 &  0.806 &  0.806   \\
\hline
\end{tabular}
\caption{Precision, recall, and f1 scores for $\gamma$-proton-helium-carbon shower images with energies $E=60$~TeV and with zenith angle $\theta_0 = 45^\circ$.}
\label{table_SM_4}
\end{table}

\section{Anomaly finder robustness checks}
\label{appendix_anomaly}
\begin{figure}[t]
    \centering
    \includegraphics[scale=0.8]{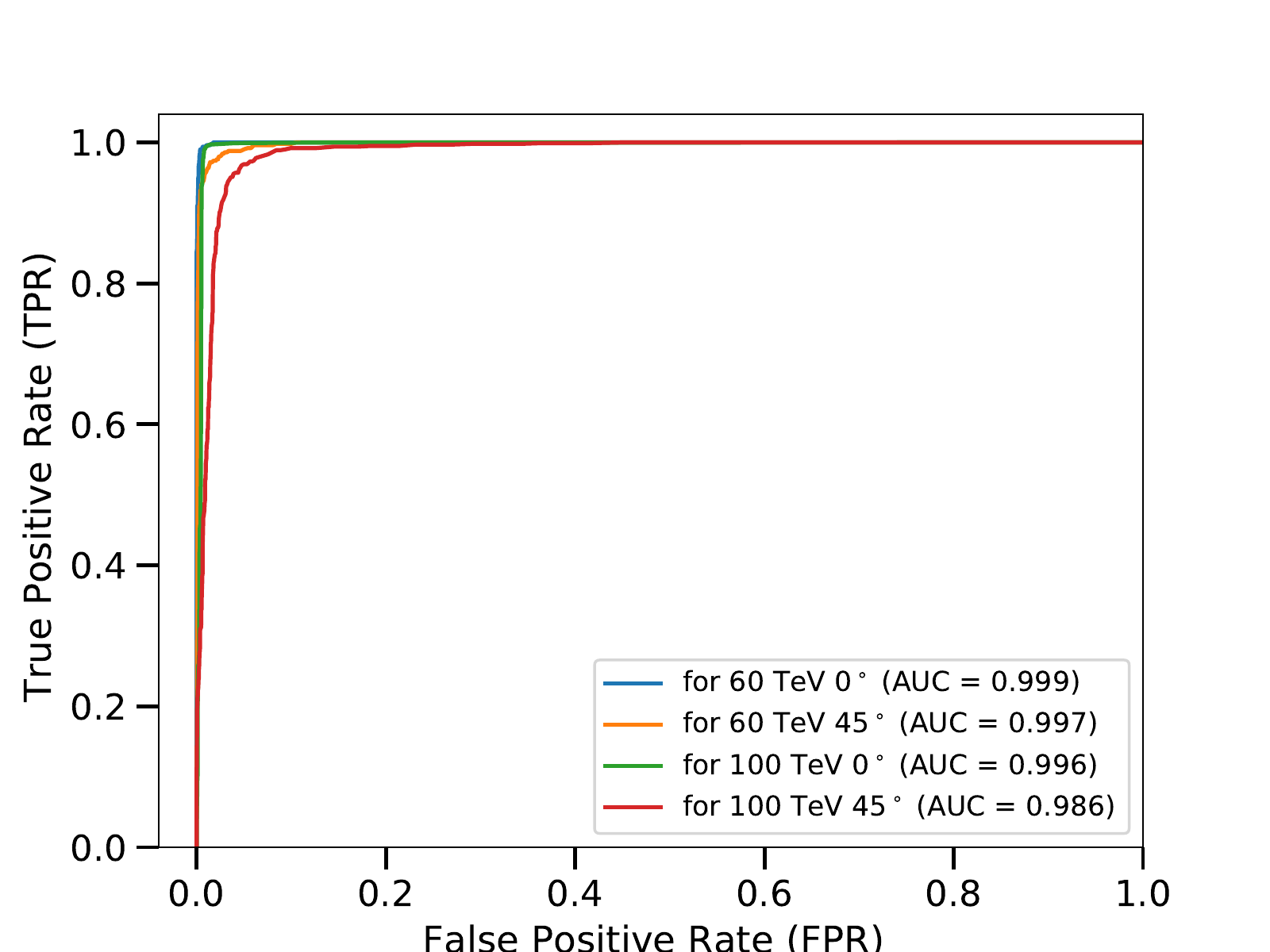}
    \caption{ROC curves for our anomaly finder for all energy and zenith angle combinations.}
    \label{fig:ROC4}
\end{figure}

We have also repeated the simulations for our autoencoder by training on Standard Model initiated cosmic rays showers with other choices of energies and zenith angles for the primary and for the $\zp$. However, for the $\zp$ testing images we only simulated 1000 images for the (100~TeV, 45$^\circ$) case and 500 images each for the (60~TeV, 0$^\circ$), and (60~TeV, 45$^\circ$) cases. This is because at lower energies, the $\zp$s yield a lower rate for triggering all four telescopes, thus a very large number of $\zp$ events need to be simulated in \texttt{CORSIKA} to obtain viable shower images.
The resulting ROC curves for all energy and angle combinations are shown in Fig.~\ref{fig:ROC4}. For a 5\% FPR we find a TPR $> 95\%$ for all energy and angle combinations.

\begin{figure}
    \centering
    \includegraphics[scale=0.73]{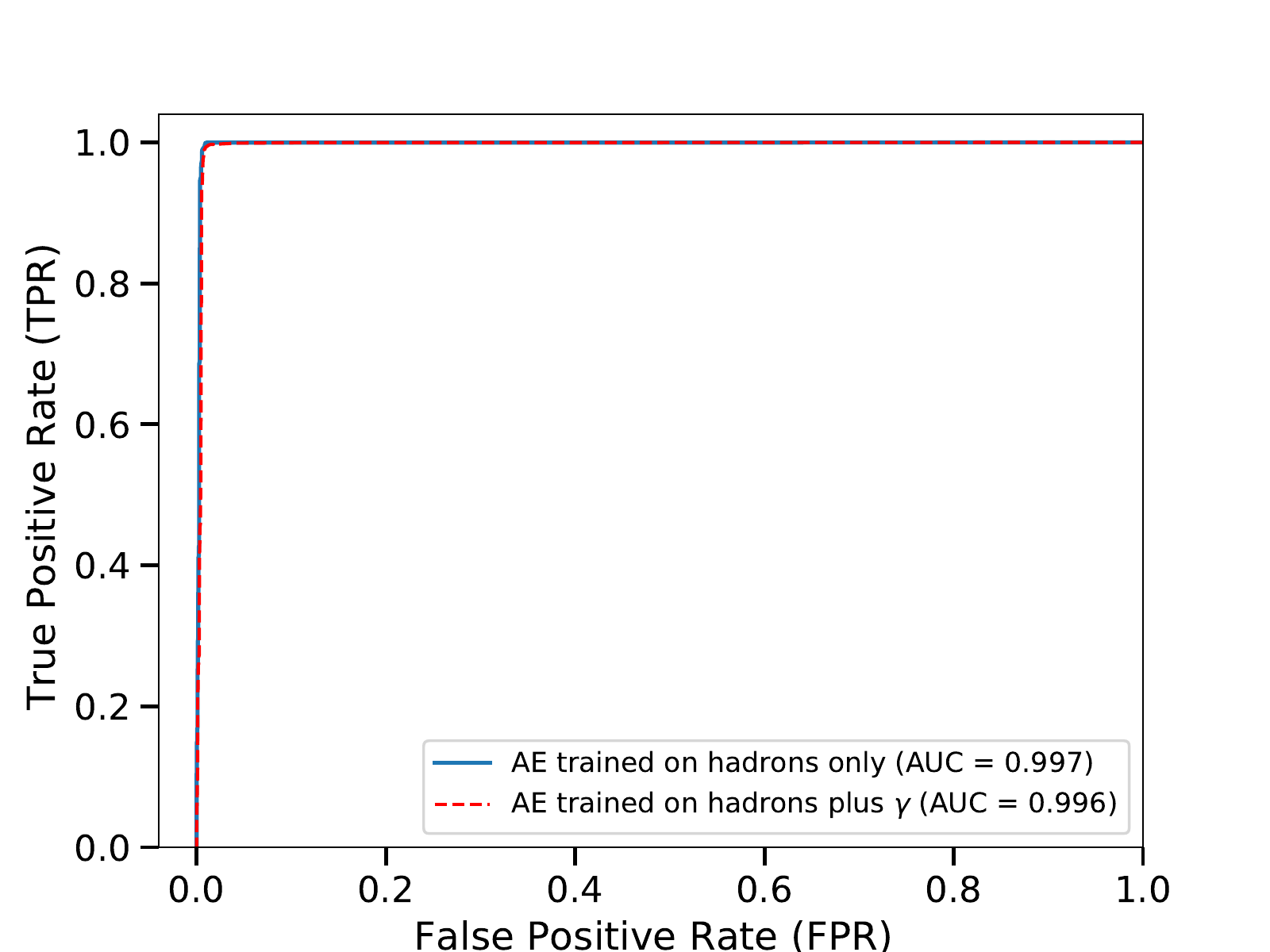}
    \caption{ROC curve for our autoencoder trained only on hadronic showers at 100 TeV and $\theta_0 = 0^\circ$. The performance is similar to that of the autoencoder trained on hadrons and gamma ray showers.}
    \label{fig:ROCHadron}
\end{figure}

We have also attempted to check the robustness of the anomaly finder when trained only on hadrons (i.e. $p$, He, and C, without including $\gamma$ ray showers), and tested on a mixed sample of hadronic and $\zp$ shower images. The resulting ROC curve for anomaly detection with primaries at 100 TeV and $\theta_0 = 0^\circ$ is shown in Fig.~\ref{fig:ROCHadron}. We see that the ROC curve is similar to that of Fig.~\ref{ROC}, which we have reproduced again here for comparison. Thus, the anomaly finder does not appear to have any trouble discriminating anomalous $\zp$ events from hadronic showers.

\bibliographystyle{JHEP.bst}
\bibliography{CosmicRay_Ref}

\end{document}